\documentclass[aps, twocolumn, groupedaddress,superscriptaddress, showpacs,pre,nofootinbib]{revtex4-1}
\usepackage{graphicx}
\usepackage[title]{appendix}
\usepackage{array}
\usepackage{setspace,transparent}
\usepackage{color}
\usepackage{afterpage}
\usepackage{fontenc,subcaption,ragged2e,amsmath}
\usepackage{mathtools,amssymb,nicefrac,tikz}
\usepackage{soul}
\usepackage[font=small,labelfont=bf]{caption}
\usepackage[utf8]{inputenc}
\usepackage[absolute,overlay]{textpos}
\usepackage{capt-of}
\usepackage{url}
\usepackage{float} 
\graphicspath{ {./figures/} }

\DeclareCaptionJustification{justified}{\justifying}
\newcommand{\RN}[1]{%
 \textup{\uppercase\expandafter{\romannumeral#1}}%
}
\DeclareMathAlphabet{\mathpzc}{OT1}{pzc}{m}{it}

\begin{document}

\title{The role of dependency range in thermally coupled superconducting networks}

\author{Nahala Yadid} 
\affiliation{Department of Physics and Jack and Pearl Resnick Institute and Institute of Nanotechnology and Advanced Materials, Bar-Ilan University, Ramat-Gan 52900, Israel}
\author{Bnaya Gross}
\affiliation{Network Science Institute, Northeastern University, Boston, MA 02115}
\affiliation{Department of Physics, Northeastern University, Boston, MA 02115}

\author{Shlomo Havlin}
\affiliation{Department of Physics and Jack and Pearl Resnick Institute and Institute of Nanotechnology and Advanced Materials, Bar-Ilan University, Ramat-Gan 52900, Israel}
\author{Nadav Shnerb}
\email[Corresponding author: ]{nadav.shnerb@gmail.com}
\affiliation{Department of Physics and Jack and Pearl Resnick Institute and Institute of Nanotechnology and Advanced Materials, Bar-Ilan University, Ramat-Gan 52900, Israel}
\date{\today}

\begin{abstract}
Interdependent networks can be highly fragile because dependency links amplify local failures into cascades that trigger global collapse. Much of the theoretical understanding comes from geometric network models, where dependency range is imposed externally and coupling strength is distance independent, and it is unclear how these insights apply to real materials, where interdependence arises from physical transport. Here we study thermally coupled amorphous superconducting networks, in which dependencies emerge from current redistribution and heat-mediated interactions governed by thermal diffusivity. Using an experimentally calibrated model of resistively shunted Josephson junctions, we show the system reproduces the three canonical transition regimes: continuous, nucleation-dominated, and mixed-order. However, because heat-driven dependency simultaneously broadens and weakens with distance, long-range thermal coupling can stabilize the superconducting phase upon heating, in contrast with its destabilizing role in geometric models. Cascades are anisotropic, with one-dimensional front propagation. These results offer an experimentally testable perspective complementing the geometric framework.
\end{abstract}

\maketitle

\section*{Introduction}

Network science has gained considerable momentum over the past decades, with applications to internet, transportation, electrical, biological, and social networks, among others~\cite{pocock2012robustness,mitchell1974social,smith2008social,yeang2004physical,newman2003structure,rucker2012network,newman2011structure,cohen2010complex}. One of the most important aspects of the analysis of networks has to do with their stability against disturbances, such as removal of nodes or links, in particular with regard to the likelihood that local damage will propagate and percolate through the network~\cite{essam1980percolation, shante1971introduction,li2012cascading}. Networks, by themselves, are geometric structures defined entirely by their nodes and the links between them. Naturally, when we come to discuss and consider real-world physical systems (PS), it is necessary to examine what is similar and what differs between an abstract, geometric network model and an actual physical system.

During the last two decades, many studies have focused on the dynamics of interdependent networks (IN)~\cite{buldyrev2010catastrophic,helbing2013globally,klosik2017interdependent,scheffer2001catastrophic,bashan2013extreme,radicchi2017redundant}. The simplest scenario of interdependency involves two networks, where the nodes in one network are connected to nodes in another network by {\it dependency links}. An example is an electrical grid and the network of servers that control its activity: the grid’s activities depend on the servers, while the servers themselves need electricity to function~\cite{rinaldi2001identifying,laprie2007modelling}. 

The main implication of interdependency is the emergence of cascading failures~\cite{rocha2018cascading,lawrence2020cascading}. This feature has been studied in depth in models of geometric networks~\cite{bashan2013extreme}, not only for the case of two networks with interdependence between them, but also for a single network combining connectivity links and dependency links~\cite{sallem2025interdependentnetworkstwointeractionsphysical}. In particular, it has been found that the characteristic range of the dependency links plays a critical role for the robustness of the system~\cite{li2012cascading}. When this range is very short, only short cascading failures develop and the transition is that of ordinary percolation, that is a continuous second-order; when the range is intermediate, there is nucleation and spreading of the failure, as in first-order phase transitions. Finally when the range is sufficiently large, a mixed-order transition appears, characterized by discontinuous jump and critical scaling above the transition~\cite{li2012cascading}. 

Recent work has effectively demonstrated that some key features of interdependent network theory are experimentally valid using a controllable and physically straightforward experimental system, which involves thermally connected layers of amorphous superconductors~\cite{bonamassa2023interdependent,gross2025random}. In this system, a percolating cluster of superconducting (SC) links is required for current flow, while the Joule heating due to failed (Normal) resistors affects other links and provides long-range dependency mechanism. The experimental results showed excellent agreement with the numerical results obtained from spatial simulations based on the RSJJ (Resistively Shunted Josephson Junction) model.

Despite the system’s microscopic complexity and the many intertwined physical effects, its failure dynamics admit a simple organizing framework rooted in geometric interdependent-network theory. 

\begin{figure*}
	\centering
 \includegraphics[width=1\linewidth]{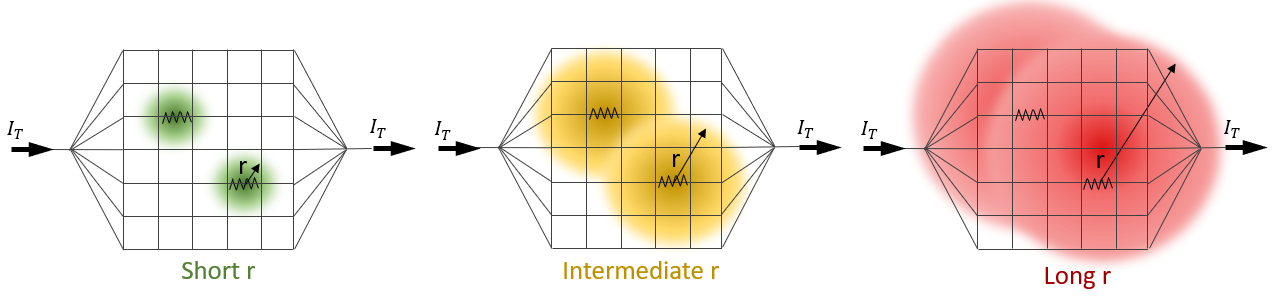}
 \caption{{\bf A spatial network of amorphous superconducting segments.} Current $I_T$ flows from left to right through the network. Each segment $(i,j)$ of the 2d lattice represents a link (Resistively Shunted Josephson Junction, RSJJ), characterized by its own critical current value, $I^c_{i,j}(T)$. As temperature varies, a few segments may change their state from superconducting to normal and vice versa. Zigzag lines denote resistive (normal) segments that generate heat, which subsequently diffuses with diffusion constant $D$. The radius $r=\sqrt{Dt}$ (colored circles) is the typical heating range around each heat source, where $t$ is governed by the response rate of the system. This figure illustrates the single-layer SC setup. In Supplementary Note~4 (Figs.~S8-S10), we present similar results for a two-layer configuration, where heat can flow between the layers. }
 \label{fig:Ilustration}
\end{figure*}

Motivated by these findings, here we set out to theoretically analyze the system of amorphous superconductors with thermal dependency of {\it limited range}, and compare it to the geometric network studied in Ref.~\cite{li2012cascading}, and to identify the similarities and differences between the two systems, the physical and geometrical systems. We find that although the superconducting networks (SN) admits many of the characteristics of the geometric network model, it differs from it by three important features (see Fig.~\ref{fig:Ilustration}), yielding phenomena that could be tested in future experiments: 

\begin{itemize}

\item Because current flow introduces a preferred direction, the dynamics become anisotropic, in contrast to the geometric model. Consequently, one must distinguish between longitudinal avalanches, which propagate along the direction of the current, and transverse avalanches, which expand perpendicular to it.

\item In addition to its connectivity links and the heat-mediated dependency links, the physical network also exhibits a third type of effective “link” arising from the redistribution of current load. When a segment becomes normal and its resistance increases sharply, the current is diverted into adjacent segments. This redistribution of current can, in turn, drive those neighboring segments toward failure, providing an additional pathway for cascading processes.

\item The strength of the dependency links in our SN system is range-dependent. In geometric models the strength of a dependency link does not depend on its length. Heat-mediated dependencies behave differently: when the thermal diffusivity is high, the same amount of heat spreads over longer distances, so its influence on any particular neighboring site is weaker. Conversely, when the thermal diffusivity is low, heat remains localized, i.e., it does not travel far, but its concentration near the source is higher, making its effect on nearby sites stronger. This coupling between range and strength does not exist in the geometric models, where dependency links are typically treated as distance-independent within range $r$.

\end{itemize} 

In this work we show that, despite the continuous and distance-dependent nature of heat-mediated dependency, the system still exhibits the three transition regimes found in geometric dependency models. For short dependency ranges (low thermal diffusivity) the transition is continuous, for intermediate ranges the collapse proceeds through spontaneous nucleation and subsequent expansion, and for sufficiently large ranges a mixed-order transition emerges, combining a macroscopic jump with critical scaling above the transition. Besides, the critical temperature exhibits, as in geometric IN, a non-monotonic dependence on the dependency interaction length, here, in PS, controlled by the heat-diffusion constant.

On the other hand, we identify, due to the above features in PS, several phenomena that are absent from the canonical isotropic geometric model used here as a benchmark. First, transverse and longitudinal cascades behave differently, making the nucleation process strongly {\it anisotropic}, an effect that plays a central role in shaping nucleation dynamics. Second, long-range dependency links can {\it stabilize} the system, in contrast to their typically destabilizing role in geometric models. This stabilization arises from the interplay between distance and coupling strength in heat-mediated interactions, as well as from the current-redistribution mechanism. We emphasize that these findings complement the geometric framework rather than replace it: they probe a physically motivated setting in which some of its standard assumptions are relaxed. 

\section*{Model system and its implementation}

Motivated by the experiment of Ref.~\cite{bonamassa2023interdependent} and subsequent works, we consider a square ($L \times L$) lattice of amorphous superconducting material (Fig.~\ref{fig:Ilustration}), where segments are considered as RSJJs. To simplify the analysis, we implement in our simulations the approximate current-voltage characteristic used in Ref.~\cite{bonamassa2023interdependent}, as explained in the Methods. Accordingly, the resistance $R_{ij}$ of each segment $(i,j)$ may be negligibly small (when the superconducting channel is dominant), high (normal), or intermediate, depending on the ratio between $I_{ij}$, the current through this segment, and $I_{ij}^c$, the maximal superconducting current (see Methods). The total current in the system, $I_T$, is imposed externally, so that a local increase in resistance forces the current to redistribute into other segments. 

Since $I_c$ depends on temperature (see Methods), as the system heats up or cools down, segments may change their state, from normal to intermediate, to superconducting and vice versa. High-resistance segments generate significantly more heat per unit current through Joule heating than their superconducting counterparts. This heat then spreads to neighboring regions, influencing the $I_c$ values of adjacent segments. Consequently, the transition of some segments from one state to another can trigger similar transitions in nearby segments, potentially leading to local or even global avalanches. 

As the temperature increases, more segments transition from the superconducting state to the normal state. However, since the material is amorphous, each segment $(i,j)$ undergoes this transition at a different temperature $T^c_{i,j}$ (see Methods).
If Joule heating is neglected, as in the case of perfect coupling between the network and the heat bath, raising the temperature will cause segments at random locations in the network to switch states. The resulting behavior corresponds to that described by the classical percolation model of a resistor network~\cite{kirkpatrick1973percolation}, exhibiting a continuous phase transition.

In IN, such as the geometric models studied in~\cite{buldyrev2010catastrophic,bashan2013extreme}, two types of links are present: connectivity links, as in classical networks, and dependency links, which transmit the failure of one node to another. In these models the connectivity nodes are typically irreversible and the dependency links act in a single direction: they propagate failure but not recovery. In our system, in contrast, both aspects differ. The connectivity links are reversible: a segment can not only fail (turn normal) but also recover (return to the superconducting state), depending on the local temperature. Correspondingly, the heat-mediated dependency links can act in both directions: they can propagate not only failure but also recovery. As a consequence, whereas the geometric cascade proceeds in a single direction, here the same dependency mechanism drives cascading processes in both temperature-sweep directions. Upon heating, the failure of a superconducting segment turns it normal, so that either the current is diverted around it or it dissipates more heat; in both cases neighboring segments are triggered to switch to the normal state. Upon cooling, the reverse process propagates: a segment that switches to the superconducting state stops dissipating heat, cooling its neighbors and promoting their transition as well. In both directions the same physics of interdependent networks is at play.

In percolation of geometric networks, an attack on the network involves a random removal of each node with probability $1-p$, and the network collapses at $p_c$, below which its giant component, $P_\infty(p)$, is no longer macroscopic. In the absence of dependency links, the transition is continuous (second order) and gradual. When dependency is introduced~\cite{li2012cascading} the range of dependency links, $r$, governs both $p_c$ and the nature of the critical phenomena. When the dependency is short-range, it has little to no effect on these properties which results in a continuous, second-order transition. In contrast, long-range dependency gives rise to a mixed-order phase transition characterized by a discontinuous jump, critical exponents and a hysteresis loop. As the system approaches the transition point at $p_c$, the order parameter scales as $(p - p_c)^{\beta}$, with $\beta = 1/2$~\cite{parshani2010interdependent}. For intermediate interaction ranges, weak regions nucleate as localized failure and propagates through the network, leading to an abrupt collapse without the preceding power-law behavior. The critical point $p_c$ is a non-monotonic function of $r$, and the system is most sensitive (higher $p_c$) when $r$ resonates with the typical radius of weak regions, correlation length, in the infinite cluster~\cite{li2012cascading}.

In these geometric models, each dependency link connects two randomly chosen nodes separated by a distance smaller than the prescribed $r$, which is the maximal range of dependencies. In our physical SN, this is not the case. Heat is transferred diffusively through the system, affecting all segments at a certain range, and the randomness has to do with the different heat-sensitivity of each segment of such an amorphous material. We accounted for this effect by simulating heat transfer explicitly, using spectral method to solve the heat diffusion equation (see Methods).
In analogy with the dependency range $r$ in the geometric models we define, in our physical model $r \equiv \sqrt{D t}$, where $D$ is the diffusion coefficient and $t$ represents the time scale determined by the system's response rate to the propagating heat (See Supplementary Fig.~S7). When $D$ is large, the heating due to a given normal segment is uniform and weak, whereas if $D$ is small, the heating is localized and strong, as illustrated in Fig.~\ref{fig:Ilustration}. 

Most studies of IN have focused on two distinct networks coupled through dependency links~\cite{buldyrev2010catastrophic,bashan2013extreme}. However, it has been shown that the same results can be obtained in a single network that contains both connectivity and dependency links~\cite{parshani2011critical}. This equivalence has also been experimentally confirmed in the amorphous superconducting networks discussed here~\cite{sallem2025interdependentnetworkstwointeractionsphysical,li2012cascading}. In what follows we implement simulations of both cases, one layer and two layers, depending on which of them illustrates better the effect we are discussing, but the canonical transition regimes and the principal cascade-dynamical features appear in both architectures; the stabilization result shown here is demonstrated for the two-layer heating protocol. 

\begin{figure*}
	\centering
	\includegraphics[width=\textwidth]{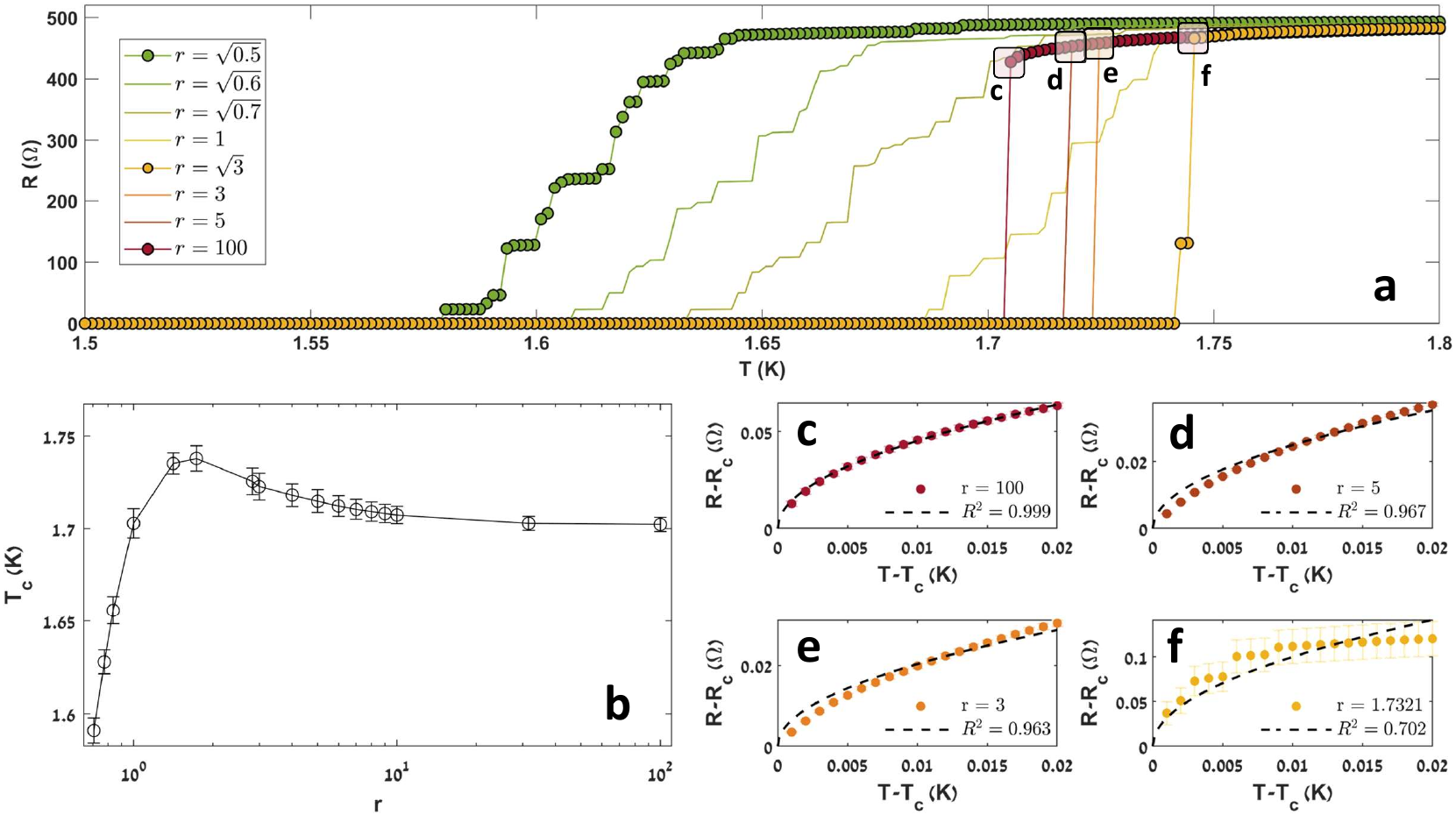}
	\caption{\textbf{Dependence of the superconducting transition on interaction range. }({\bf a}) Resistance $R(\Omega)$ vs temperature $T$ through the cooling process. The total resistance of the network is high at elevated temperatures, when all segments are normal. As the temperature decreases, some segments transition to the superconducting state, drawing current away from their neighbors and decreasing their temperature, thus increasing the likelihood that those segments will also enter the SC state. As the process progresses, more segments transition to the superconducting state, until eventually the total resistance drops to zero. The range of dependencies in the system is governed primarily by thermal diffusivity through the parameter \(r\) (see Fig.~1). By implementing an iterative solution of the Kirchhoff equations we plotted several resistance-temperature curves illustrating the transition for different values of \(r\). As in Fig.~\ref{fig:Ilustration}, we distinguish between short (green), intermediate (yellow), and long (red) $r$-values. For an extended version of this panel see Supplementary Figure S11. Each curve in panel (a) shows a single representative realization. The small step-like features these curves exhibit are finite-size effects of individual realizations. They are distinct from the macroscopic transition (see Supplementary Note~1). The highlighted markers (see legend) indicate three representative cases, one from each regime. As in geometric networks~\cite{li2012cascading,xue2024nucleation}, the system transitions from ordinary continuous transition percolation at low values of \(r\), to nucleation at intermediate values, and to a mixed transition at high values. (\textbf{b}) The critical temperature $T_c$ is plotted against $r$. $T_c$ is identified using the resistance-threshold criterion (the temperature at which the normalized resistance crosses $R=0.1$ of the normal-state value), and each point is the mean over $n=100$ independent disorder realizations; error bars denote one standard deviation across realizations. This definition agrees, within the scatter, with three other operationally distinct criteria (maximal slope, appearance of a spanning superconducting cluster during cooling, and peak in convergence time), as shown in Supplementary Notes~1 and~2.
 Again, similar to geometric networks~\cite{li2012cascading}, the system is most sensitive at intermediate values of \(r\), which promotes rapid propagation of alternative phase nuclei. For values of \(r\) that are very small, nodes that switch state cool their immediate surroundings, which are themselves part of the nucleation seed, so heat spreading has only a minor effect. For values of \(r\) that are very large, the heat spreads broadly and becomes ineffective for nucleation.
(\textbf{c-f}) As seen in panel (a), a jump in the order parameter (the resistance) as a function of temperature appears both in the mixed-order transition, where it arises from global cascading failure, and 
in cases where the transition is driven by nucleation, as in standard first-order transitions. The curves in panels (c-f) are ensemble averages over the same $n=100$ realizations, with error bars defined as in panel (b). What distinguishes between nucleation transitions and mixed-order transitions is the critical exponent \(\beta\) right above the jump point. For large $r$ values, the data are consistent with \(\beta = 1/2\) as expected (Eq. \ref{eq0}), whereas at intermediate values of \(r\), where the transition occurs via nucleation, no clean power law is observed. We caution that a precise quantitative determination of $\beta$ is limited by the uncertainty in the location of $T_c$, to which the fit is sensitive; the $\beta=1/2$ fits shown here (dashed lines) are therefore intended as qualitative support for the transition-type classification rather than as a precise measurement of the exponent.
All these results were obtained for a square \(L=100\) network, subject to \( I_T = 500\, \mu\mathrm{A} \). The dependence on system size $L$ is examined in Supplementary Figure S4. Heat is converted to temperature using a heat-retention coefficient \(\gamma=10^8 ~\mathrm{W^{-1}\,K}\) (see Methods). The coefficient of determination, $R^2$, indicates how well the $A\sqrt{T - T_c}$ fit describes the averaged collapse data, with values near $1$ denoting better agreement.}
\label{fig:2_1_mixed_Tc}	
\end{figure*}

\section*{Results}

\subsection*{Transition regimes and the critical temperature}

As explained above, our goal here is to examine how, and to what extent, the effects demonstrated in geometric dependency networks also appear in simulations of the physical SC system, to characterize the differences in the outcomes, and to attribute them to the physical mechanisms that arise in our SC model.

We begin with the features that are shared between the systems. Fig.~\ref{fig:2_1_mixed_Tc} shows that, in agreement with geometric interdependent networks, increasing the dependency range $r$ changes the nature of the phase transition. In the present example we consider cooling of the SC-N (superconducting--normal) network, so the total resistance drops to zero (or very close to zero) at the transition.

Identifying the transition in this system requires care, since the appropriate order parameter is not a priori obvious. We therefore located $T_c$ using four operationally distinct criteria: the resistance threshold, the maximal slope of $R(T)$, the appearance or loss of a spanning superconducting cluster (depending on the sweep direction), and the peak in the number of iterations required for convergence. These yielded mutually consistent estimates of $T_c$ (Supplementary Note~1). Moreover, the sample-to-sample fluctuations of $T_c$, quantified by the standard deviation of its distribution over realizations, decrease with system size $L$ (Supplementary Fig.~S4); the observed decrease is consistent with a sharpening transition over the accessible sizes under the fixed-total-current protocol. Together, these results establish that the transition is well defined and not an artifact of a single realization or of a particular measurement convention.

The critical temperature depends non-monotonically on the dependency range $r$, peaking at an intermediate value (Fig.~\ref{fig:2_1_mixed_Tc}(b)), a behavior familiar from geometric interdependent networks. Li \textit{et al.}~\cite{li2012cascading} found that the network becomes most vulnerable (and collapses earlier) when the dependency range matches the characteristic size of the holes (correlation length) created after the initial attack. If the dependency range is larger, cascades originating from nodes that failed inside the hole tend to affect nodes far from the hole's boundary, and therefore do not contribute effectively to the propagation of the nucleation front. Conversely, if the dependency range is too small, most of the cascade is wasted on nodes that are already within the damaged region. As Fig.~\ref{fig:2_1_mixed_Tc}(b) indicates, the same phenomenon appears in the amorphous superconductor system, as the critical temperature peaks (upon cooling) at a certain intermediate value of $r$, likely reflecting scale matching between $r$ and the size of natural holes.

Guided by this analogy, we expect the physical network to display the same succession of transition types as the range is varied: a continuous, percolation-like transition at small $r$, a nucleation-dominated transition at intermediate $r$, and a mixed-order transition at large $r$. As we now show, each of these regimes leaves a distinct signature in our data.

At small $r$ the transition is continuous: the resistance decreases gradually over a broad temperature interval, as in ordinary percolation. As $r$ increases this gradual behavior is progressively lost and the transition sharpens into an abrupt jump, as seen directly in the resistance curves of Fig.~\ref{fig:2_1_mixed_Tc}(a), which evolve from smooth and broad at small $r$ to sharp and step-like at larger $r$.

\begin{figure*}
	\centering

	\includegraphics[width=\textwidth,
 trim=20 150 90 130, 
 clip] {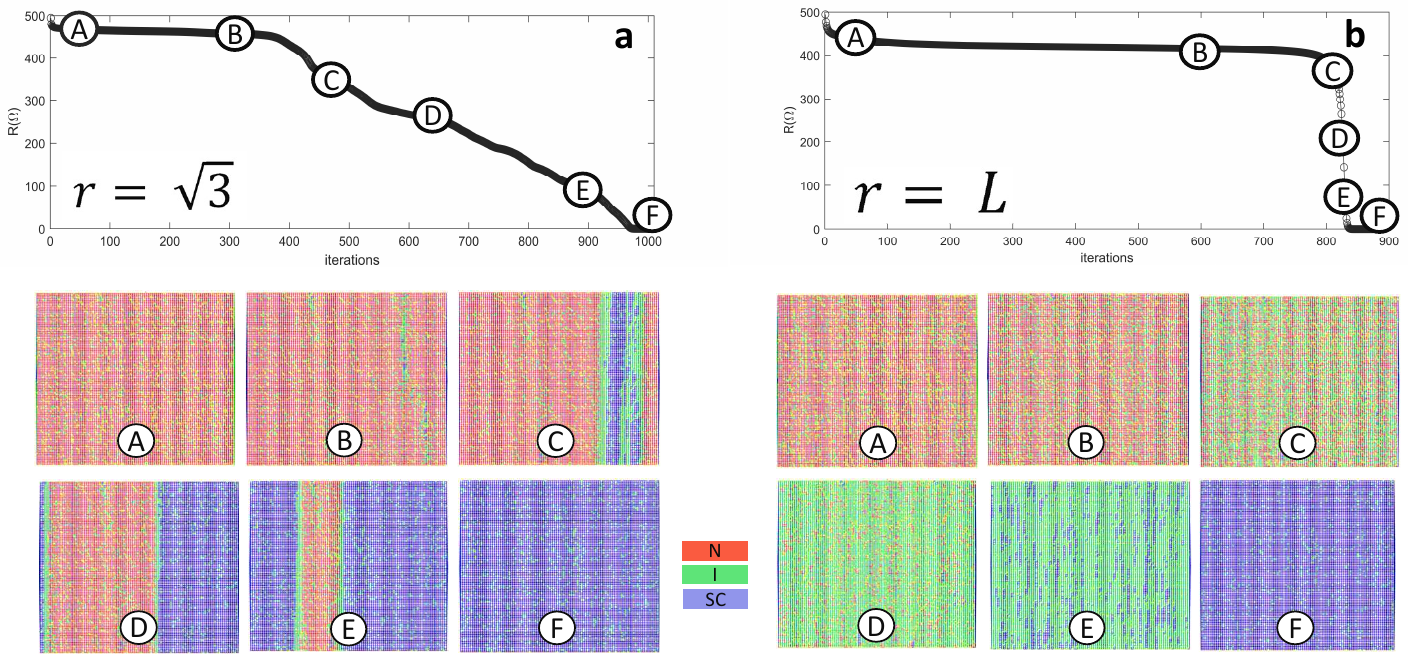}
	\caption{{\bf Spontaneous cascading dynamics of the collapse at the critical temperature.} 
The global resistance of a square network ($L = 100$) is plotted 
against simulation ``time" (number of iterations), during the abrupt collapse, after cooling from the conducting state at $T = 3 K$ to the critical temperature. Panels ({\bf a}) and ({\bf b}) 
correspond to two of the dependency ranges considered in 
Fig.~\ref{fig:2_1_mixed_Tc}. Snapshots of the system are shown at the points (A--F) marked along each resistance curve. When the dependency length is intermediate (panel (a), $r=\sqrt{3}$), one 
observes the typical behavior of a first-order transition: islands of 
superconducting segments (blue) emerge within the normal network (red) and begin 
to expand. Green links represent an intermediate state between the normal and 
superconducting phases (see~\cite{bonamassa2023interdependent} and the Methods 
section below). The anisotropy of the system manifests itself in the fact that 
nuclei expand first in the transverse direction and only later in the 
longitudinal one. Consequently, the total resistance in panel~({\bf a}) decreases 
approximately linearly with time, in contrast to what is observed in geometric 
networks~\cite{li2012cascading} ($T_{c,r=\sqrt3} = 1.74 K$). On the other hand, in panel~({\bf b}), where $r = L$ ($T_{c,r=100} = 1.701 K$), we reproduce the qualitative 
behavior observed experimentally in~\cite{bonamassa2023interdependent}: a long 
plateau during which the resistance remains nearly constant, followed by a sharp 
drop. The plateau reflects a non-local and non-nucleation cascading-failure process in which each segment that becomes superconducting pulls, on average, one randomly located additional segment into the superconducting phase, corresponding to a critical branching process that evolves very slowly~\cite{dilmoney2025dynamics}. The snapshots in (b) show that, although some anisotropy is still discernible, it is much weaker than in (a), the $r=\sqrt{3}$ case. This indicates that the dominant mechanism here arises from 
heating/cooling rather than from current redistribution, which is identical in 
both cases. All results are for the same networks defined in Fig.~\ref{fig:2_1_mixed_Tc}, 
using the same parameters.}
	\label{fig:different_colapses}	
\end{figure*}

The boundary between the nucleation regime and the mixed-order regime is identified by the appearance of critical scaling above the jump, characterized by the critical exponent $\beta$, defined by
\begin{equation} \label{eq0}
 R - R_c \sim (T - T_c)^\beta
\end{equation}
where $R$ is the network’s global resistance which serves as an order parameter and $T$ is the heat-bath temperature. $R_c$ is the global resistance at the transition temperature $T_c$.
A mixed-order transition is characterized by $\beta = 1/2$~\cite{gross2025random}, and for large $r$ our data are consistent with this square-root behavior (Fig.~\ref{fig:2_1_mixed_Tc}(c)). In contrast, in the nucleation regime the transition is governed by metastability and the spontaneous abrupt collapse without any curvature representing the growth of a finite nucleus, rather than by a critical fixed point. As a result, no power-law scaling develops near and above $T_c$ and no square-root behavior is observed (Fig.~\ref{fig:2_1_mixed_Tc}(d)-(f)). We stress that this classification does not rest on a precise determination of the exponent alone: the two regimes are distinguished by the presence or absence of critical scaling above the jump, together with the markedly different spatial and dynamical signatures of the collapse discussed below (Fig.~\ref{fig:different_colapses}).

This distinction also has a clear spatial and dynamical signature: in the nucleation regime (intermediate $r$) defective clusters nucleate locally and then propagate through the system, whereas in the mixed-order regime (large $r$) the collapse proceeds through a pronounced plateau followed by random, spatially delocalized failures, as shown by the collapse dynamics and snapshots in Fig.~\ref{fig:different_colapses}(\textbf{a}) and (\textbf{b}), respectively.

\subsection*{Anisotropic cascade dynamics}

We now turn to the aspects in which the physical system significantly {\it differs} from geometric dependency networks. First, as seen in Fig.~\ref{fig:different_colapses}(a), the physical system exhibits broken isotropy, and it is necessary to distinguish between longitudinal failure cascades (along the general direction of the current) and transverse ones. This distinction is most pronounced when the phase transition occurs via nucleation. In this case, we observe a transverse expansion of the opposite phase: in each cascade event, a single entire column of links switches from the normal state to the superconducting state. This is followed by a longitudinal expansion, in which adjacent columns also become superconducting, until the resistance drops to zero.

Because the propagation of the superconducting phase is predominantly one-dimensional, in case of anisotropic nucleation the resistance decreases linearly in time. This is in contrast with geometric two-dimensional networks, where the order parameter (the size of the infinite cluster) increases quadratically in time~\cite{xue2026mechanistic}. 

In the case of a long dependency range ($r = L$, with $L$ the system size), the anisotropy is of course less pronounced, yet symmetry breaking can still be clearly identified in snapshots of the system, see the snapshots in Fig.~\ref{fig:different_colapses}(b).

An additional insight that can be extracted from Fig.~\ref{fig:different_colapses} concerns the two types of dependency present in the system. As noted above, in addition to heat-mediated dependency, the system also exhibits dependency through current redistribution: when a segment becomes normal and its resistance increases, the current tends to reroute through neighboring segments; conversely, when a segment becomes superconducting, it pulls current away from adjacent sites. Current redistribution has been extensively studied in schematic models such as the random fuse model~\cite{shekhawat2013damage}.

In our context, we can say that a segment that switches to the competing phase tends to induce neighboring segments to switch as well, through two distinct mechanisms: by modifying the current flowing through them, or by altering their temperature. The heat-mediated dependency is governed by thermal diffusivity and can therefore be manipulated both experimentally and in the simulations. The dependency induced by current redistribution, in contrast, is dictated by the instantaneous network configuration through Kirchhoff’s laws.

Fig.~\ref{fig:different_colapses} indicates that, at least for the system considered here, in the case of an infinite heat-dependency range no nucleation is observed. This implies that nucleation arises primarily from the heat mechanism, or from some interplay between thermal diffusivity and current redistribution. Since current redistribution already makes neighboring segments more susceptible to switching, it is reasonable to expect that heating or cooling these neighboring segments is more effective than modifying the temperature of a random segment in the network. Consequently, decreasing $r$ enhances the nucleation effect in two ways: it increases the amount of heat delivered per segment, because the Gaussian profile becomes narrower (See Supplementary Fig.~S7), and it targets segments that are already more prone to switching due to the current redistribution effect.

\begin{figure}
	\centering
	{\includegraphics[width=\columnwidth]{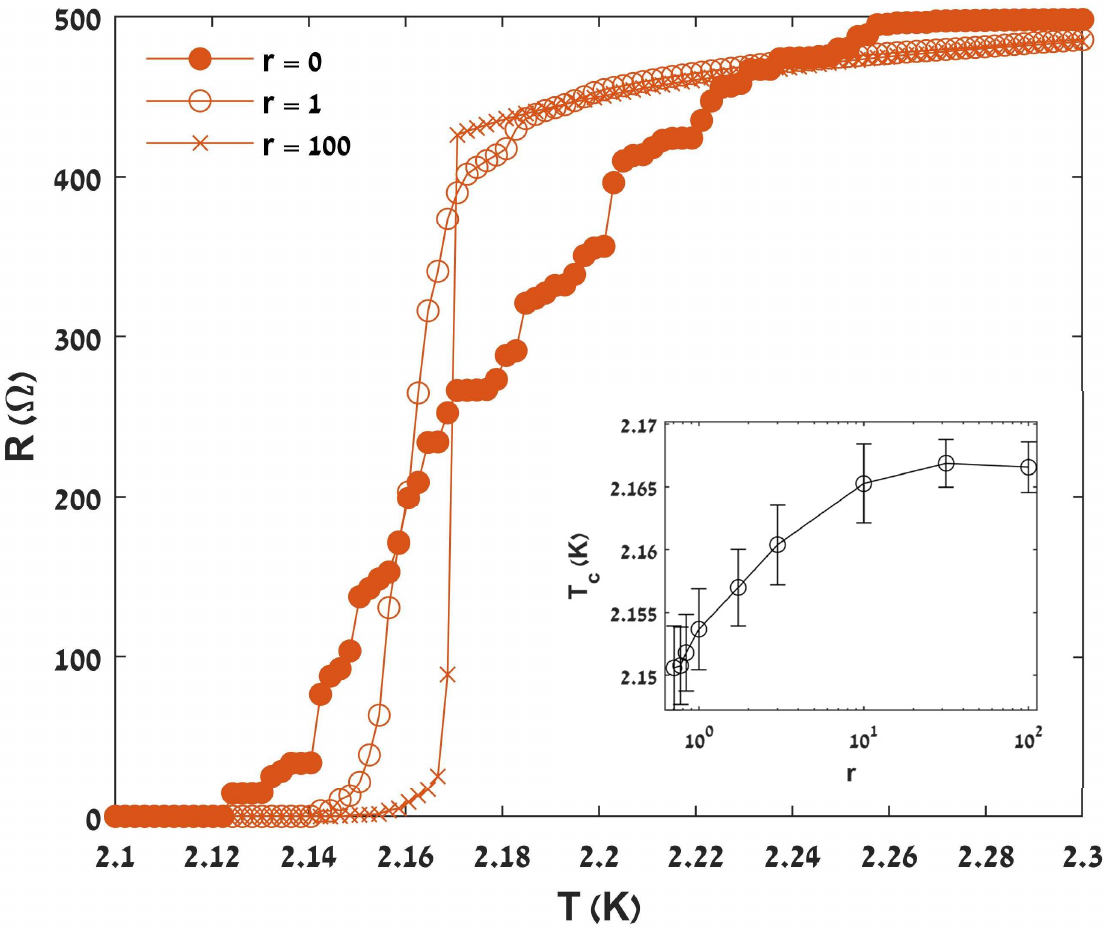}}
	\caption{ {\bf Resistance vs. temperature through heating}, for $r=0$ (full circles), $r=1$ (open circles), and very long range $r=100$ (crosses). The three curves in the main panel are single representative realizations, one per $r$; the small step-like features they show are finite-size effects of individual realizations. Averaging over many realizations, together with the other transition criteria and finite-size analysis, is presented in the inset and in the Supplementary Information. The network was initiated in its superconducting state (low T), and starts showing finite resistance when the giant component of SC links collapses. Interestingly, the network with long-range dependency is more stable than the network with $r=0$, and therefore its resistance stays zero at higher temperatures. As explained in the main text, this has to do with the ``weakening'' of thermal coupling as $r$ increases, and with the interplay between heat and current redistribution. Inset: the critical temperature $T_c$ (threshold definition, $R=0.1$ crossing; each point is the mean over $n=100$ independent disorder realizations and error bars denote one standard deviation across realizations) as a function of $r=\sqrt{Dt}$, showing that $T_c$ increases with the interaction range. The results were obtained for two networks of identical size and topology, with the layer-specific RSJJ parameters given in Methods, $L=100$, that are 
thermally (but not electrically) coupled, similar to those used in experiments~\cite{bonamassa2023interdependent}. 
 Parameters are $I_T = 300 \,\mu\text{A}$ and $\gamma = 4 \times 10^{3}~ \mathrm{W^{-1}\,K}$~\cite{sallem2025interdependentnetworkstwointeractionsphysical}.
}
	\label{fig4}	
\end{figure}

\subsection*{Stabilization by long-range thermal coupling}

An interesting manifestation of this effect appears in Fig.~\ref{fig4}, which compares the transition for a network with $r=0$ and $r=1$ to that of a network with $r \approx L$. In geometric IN, dependency is detrimental to system stability; thus, although the transition point depends on the dependency range in a non-monotonic manner, networks with effectively infinite range are still more fragile than zero-range networks. For the physical networks studied here, we observe the opposite trend: upon heating, the transition temperature is higher for the effectively infinite-range network, indicating that it is more stable than the zero-range network. As explained above, this stems from the fact that short-range heating locks a segment that has switched to the normal state into that state, and strongly affects neighboring segments that are already more sensitive due to the current-redistribution effect. While this effect, in the system examined here, is quite modest in magnitude, it is still significant (see inset of Figure \ref{fig4} and Supplementary Note~3), well resolved beyond the realization-to-realization scatter. As such, it serves as a proof of concept: as opposed to geometric networks, physical dependencies may increase the stability of a network rather than decreasing it.

\subsection*{Nucleation and branching dynamics}

\begin{figure*}
	\centering
	{\includegraphics[width=\textwidth,
 trim= 45 155 59 170, 
 clip]{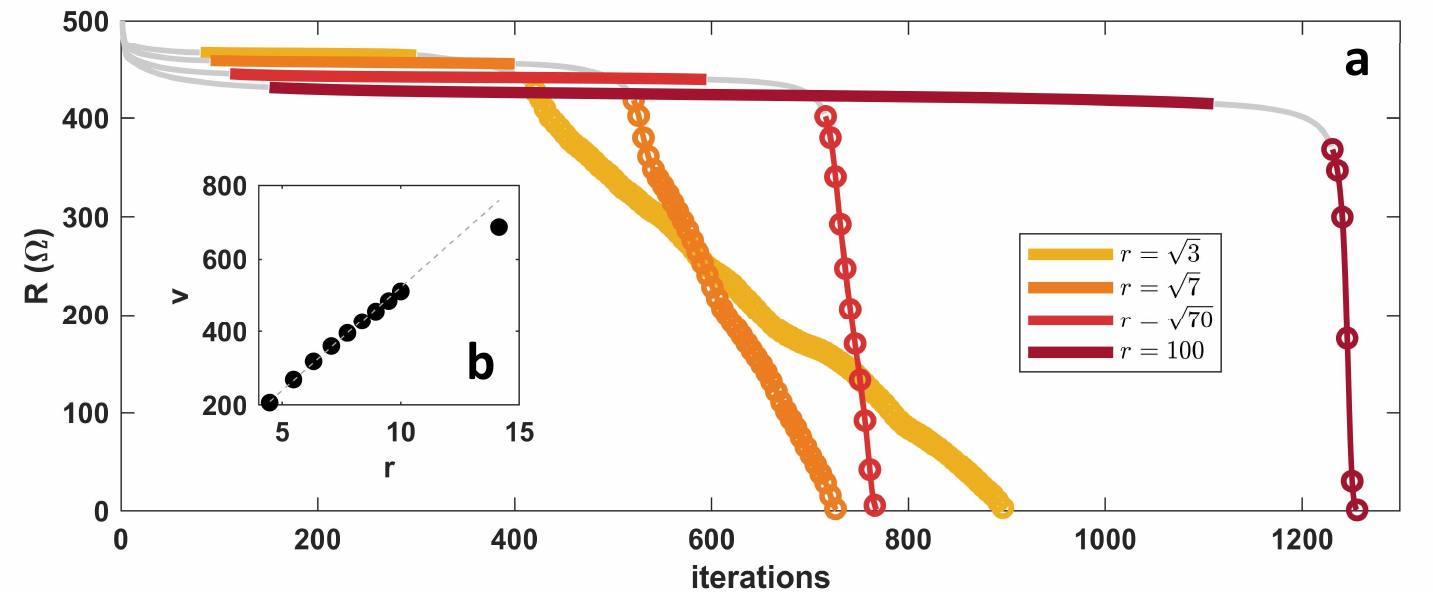}}
	\caption{\textbf{Dynamics of nucleation and branching-driven collapse. }(\textbf{a})~Network resistance as a function of time (iterations) through the abrupt transition in the nucleation (first-order transition) regime and in the random cascading case. For different values of $r$ one identifies the plateau associated with the critical branching process (thick full horizontal line part of the graphs) and the linear decrease of the resistance (thick with markers) through the nucleation. The duration of the plateau and the speed of propagation of superconducting nuclei both increase with $r$, as observed in geometric interdependent networks~\cite{xue2024nucleation}. (\textbf{b}) The mean speed of the nuclei, $v$, vs. $r$ (mean over $n=30$ independent disorder realizations), which increases approximately linearly with $r$, $v \sim r$, as also found in geometric networks~\cite{xue2024nucleation}. Minor deviations from a strictly linear trend are plausibly due to anisotropy and finite-size effects. }
	\label{fig5}	
\end{figure*}

Another difference between geometric networks and the physical network is 
provided in Fig.~\ref{fig5}. Here we show the cooling transition (resistance as a function of simulation time, as measured in iterations) in cases belonging to the abrupt-transition regime. 
A clear plateau is observed, indicating a critical branching process that persists until nucleation occurs, followed by the expansion of the nuclei. Because the expansion is predominantly one-dimensional and proceeds at a finite speed, the resistance decreases linearly in time. 
As \(r\) increases, both the duration of the plateau and the propagation speed of the nucleation seed increase: the plateau duration grows because \(r\) sets the minimal size 
of a seed that can propagate, while the propagation speed increases with $r$, rising strongly as the mixed-order regime is approached. 
Similar phenomena were observed in geometric models~\cite{xue2024nucleation,xue2026mechanistic}. Corresponding dynamical analyses for the two-layer interdependent (ISN) and single-layer self-dependent (SDSN) configurations are presented in Supplementary Note~4 (Figs.~S10 and S13).

\section*{Discussion}

In this work we have investigated theoretically via simulations, how the range of heat-mediated dependency interactions controls the stability, and the cascading failure dynamics, of an amorphous superconducting network. This framework provides a concrete physical realization of interdependent networks in which dependency links emerge from Joule heating rather than being imposed ad hoc, and it allows us to study the phenomenology of different phase transitions controlled by dependency heating range in a setting where interactions are anisotropic, continuous in space, and distance dependent. Our results call for relatively simple experiments to test these theoretical predictions.

Using an RSJJ-based numerical model, calibrated to
recent experiments on thermally coupled amorphous Indium-Oxide networks~\cite{bonamassa2023interdependent}, we analyzed the similarities and differences between abstract geometric networks such as those studied in~\cite{bashan2013extreme,li2012cascading,xue2024nucleation,xue2026mechanistic} and our physical system. We also improve our understanding of the interplay between two types of dependency: current redistribution and heat diffusive transport.

A key takeaway message is that, despite the microscopic complexity of the superconducting network and the many intertwined physical mechanisms, its abrupt transition cascading dynamics is organized by the simple principles suggested in geometrically interdependent networks.

In particular, despite these additional physical ingredients - anisotropy, current redistribution, distance dependence - many of the key features previously identified in geometric dependency networks reappear in our superconducting system. By tuning the effective dependency interaction length scale $r \equiv \sqrt{Dt}$, which is set by the thermal diffusivity, we recover the same three canonical regimes as a function of interaction range. For small $r$, the transition is continuous and closely resembles ordinary continuous percolation transition, with no evidence of cascading failures. For intermediate $r$, the breakdown proceeds via nucleation: finite clusters of
the competing phase emerge and subsequently spread through the system, giving rise to an abrupt jump in the order parameter without critical scaling above the transition. For sufficiently large $r$, the system exhibits a mixed-order transition, characterized by a macroscopic jump in the resistance followed by a critical regime in which the excess resistance is consistent with the expected square-root behavior, $R-R_{\mathrm{c}} \sim (T-T_{\mathrm{c}})^{1/2}$. In addition, the critical temperature displays a non-monotonic dependence on $r$, with a maximum at intermediate ranges where the typical interaction distance resonates with the characteristic size of nuclei, in direct analogy with geometric models of spatially embedded interdependent networks~\cite{li2012cascading}.

Nevertheless, the physical mechanism differs in important and experimentally accessible ways. Accordingly, the physical realization studied here goes beyond the geometric picture in several respects.

First, the direction of the applied current breaks the lattice symmetry and renders the failure process strongly anisotropic, in contrast to geometrical networks which are fully isotropic. We note that anisotropic geometric models of spatially embedded interdependent networks do exist~\cite{perez2022cascading,vaknin2021cascading}; there, however, the anisotropy is built into the network topology, whereas in our system the lattice is isotropic and the anisotropy is drive-induced, emerging from the interplay between the applied current and the heat-mediated dependency. Accordingly, in the SN nucleation regime, cascades proceed in two stages: transverse avalanches convert entire columns of links from the normal to the superconducting state, and only afterward does the superconducting phase invade horizontally, until global connectivity is restored and the resistance vanishes. Because the propagation front is predominantly one-dimensional, the macroscopic resistance decays linearly in time after nucleation, in contrast with the quadratic decline of the order parameter (mass of the infinite cluster) in isotropic two-dimensional geometric networks. 

The second difference between physical and geometrical networks has to do with the mechanisms underlying dependency. As explained, the superconducting network supports two dependency mechanisms: heat-mediated interactions are controlled by thermal diffusivity and can be tuned experimentally by using different substrates, while current redistribution is dictated by Kirchhoff's laws. Our results indicate that nucleation is not triggered by the redistribution alone: in the limit of effectively infinite thermal diffusivity, where the heat field becomes spatially uniform, nucleation is strongly suppressed and the transition reverts to mixed-order behavior. Nucleation instead
emerges from the interplay between current redistribution, which makes neighboring segments more susceptible to switching, and localized heating, which concentrates thermal perturbations precisely on those already fragile segments. Decreasing $r$ both increases the amount of heat deposited per segment and focuses it near the source, thereby enhancing this cooperative mechanism and promoting nucleation.

A particularly striking consequence of this interplay is the stabilization of the superconducting phase by long-range dependency interactions. In purely geometric IN models, dependency links invariably reduce the robustness of the system, and networks with very long dependency ranges are more fragile than networks with very short dependency range or without dependency at all ~\cite{li2012cascading}. In our physical system, the transition upon heating shows an opposite trend: networks with long-range thermal coupling switch to the normal state at {\it higher} temperatures, i.e., they are more stable, see Fig.~\ref{fig4}. Short-range heating locks links that have turned normal into that state and strongly affects their neighbors, which are already destabilized by current redistribution. As a result, local failures amplify and propagate more easily. When the same amount of heat is spread over a larger region, its local effect is weakened, there is no positive interference between current redistribution and heat, and the system can sustain higher bath temperatures before undergoing a global phase-transition. This shows that, in physical realizations of interdependent networks, the range and strength of dependency interactions cannot be varied independently: increasing the range may simultaneously dilute the coupling and thereby enhance stability. We stress that this stabilization is an existence result, demonstrated for the specific setup and parameter regime studied here, rather than a claim of generic behavior; whether it persists in other physical realizations remains to be tested.

Finally, our results highlight both the usefulness and the limitations of abstract interdependent-network models for describing real materials. On the one hand, the superconducting system reproduces the three regimes of phase transitions, the non-monotonic sensitivity to dependency range, and the critical branching processes underlying the plateau preceding nucleation, all of which were first characterized in geometric interdependent networks~\cite{li2012cascading, xue2024nucleation}. On the other hand, anisotropy, current redistribution, and distance-dependent coupling introduce qualitatively different behavior, such as one-dimensional front propagation and the stabilization of the ordered phase by long-range interactions. In this sense, the present work builds on the geometric interdependent-network framework rather than departing from it: the geometric picture provides the organizing principles, while the physical ingredients identified here extend its scope. 

These findings suggest that controlled superconducting networks, together with related experimental platforms such as thermally coupled ferromagnetic arrays \cite{gross2024microscopic}, can serve as versatile test beds for interdependent network theory, where both the topology and the microscopic interaction mechanisms are accessible to direct control, manipulation and measurement.

Future work could explore more complex geometries, three-dimensional architectures, and disorder distributions, as well as the investigation of how the interplay between multiple physical dependency channels shapes the resilience of other spatially embedded
infrastructure and material networks~\cite{bonamassa2023interdependent}. 

Long-range dependency, generally considered a destabilizing factor, can thus act as a source of stability in a physical setting, a possibility that awaits broader experimental and theoretical tests. We hope that our results will encourage experimentalists and theoreticians to extend the abstract geometric paradigm towards the richer spectrum of dependency effects realized in physical systems.

{\bf Funding:} N.S. acknowledges support from the Israel Science Foundation (Grant No. 2435/24). S.H. thanks the support of the Israel Science Foundation (Grant No. 201/25), the Israel Ministry of Innovation, Science \& Technology (Grant No. 01017980), the Science Ministry - Smart Mobility (Grant No. 1001706769), The EU H2020 Project OMINO (Grant No. 101086321), The Israeli VATAT project on powergrid and on climate and the DIP German-Israel agency. 

\vspace{2mm}
\noindent{\bf Author contributions:} N.Y., B.G., S.H. and N.S. contributed to all aspects of this work, including the design of the research, the development of the numerical model, the analysis of the results, and the writing of the manuscript.

\vspace{2mm}
\noindent{\bf Competing interests:} The authors declare no competing interests.

\bibliographystyle{unsrt}
\bibliography{main}

\clearpage

\renewcommand{\theequation}{M\arabic{equation}}
\setcounter{equation}{0}

\section*{Methods}

Most of the results presented through this paper were obtained for the architecture presented in Fig.~\ref{fig:Ilustration} of the main text, namely a single layer with thermal self-coupling. In what follows we will provide the details of the modeling for such a system. In a few cases (especially Fig.~\ref{fig4} of the main text) we implemented two coupled networks, in the spirit of the experimental system described in \cite{bonamassa2023interdependent}. In this setup, two networks (each having the same characteristics of a single layer) are placed on top of each other with thermally conducting (but electrical insulator) medium separating them. Therefore, the heat emitted from a normal conductor is divided into two equal parts, one stays in the original network and one is delivered to the other layer. From that point on, the numerical analysis of thermal conductivity and the update of link-states proceeds according to the procedure defined for a single network. 

\subsection*{RSJJ model of disordered superconducting networks}
To characterize the N-S PT observed in the experiments, we model the network using a disordered two-dimensional lattice of Resistively Shunted Josephson junctions (RSJJs). In the limit of a large tunneling conductance ($g \gg 1$), isolated networks of RSJJs undergo a continuous N-S PT at low temperatures that are generally independent of the ratio between the Josephson $E_J$ and the Coulomb $E_C$ energies 
\cite{orr1986global,chakravarty1986onset,chakravarty1988quantum}. In this regime, the state of each junction can be characterized by the value of its normal state resistance $R_n(T)$ and by its critical current $I_c(T)$, which generally depends on the ratio between the temperature $T$ of the cryostat and the N-S activation threshold $T_c$. The latter quantities satisfy the Ambegaokar–Baratoff relation \cite{ambegaokar1963tunneling} $I_c(T) R_n = \frac{\pi}{2e} \Delta(T) \tanh(\Delta(T) /2 k_B T)$, where the energy gap, $\Delta(T)$, follows the Bardeen–Cooper–Schrieffer mean-field spectral relation $2\Delta(T) \approx \alpha k_B T_c$ with $\alpha \approx 3.53$, $k_B$ is the Boltzmann constant, and $e$ is the elementary charge. However, the InO networks fabricated in the present work have bulk N-S thresholds large enough to ensure that junctions rarely undergo a metal–insulator transition. In light of this, we consider a model of RSJJ with only three electronic states: superconducting (SC), intermediate (IM), and normal metal (N), defined according to the Josephson I–V characteristic \cite{josephson1962possible}. Hence, the junction’s resistance is defined piecewise as:

\begin{equation}
R_{ij}=\begin{cases}
	R_{\epsilon}, & \text{if $V_{ij} < R_{\epsilon} I_{ij}^c(T)$~(SC)}\\
 R_n, & \text{if $V_{ij} > R_n I_{ij}^c(T)$~(N)}\\
 V_{ij}/I_{ij}^c(T), & \text{otherwise~(IM)}
 \end{cases}
 \label{eq:JJ_characteristics}
\end{equation}

where $R_{\epsilon}$ is the resistance in the SC state ($R_{\epsilon} = 10^{-5} \Omega$ in simulations), and $V_{ij}$ is the potential drop measured at the ends of the junction. For critical currents, we used a local generalization of the de Gennes relation \cite{de1976relation}: 

\begin{equation}
 I_{ij}^c(T_{ij}) = 
 \begin{cases}
 I_{ij}^c(0)\,(1 - T_{ij}/ T_{ij}^c)^2, & T_{ij} < T_{ij}^c,\\
 0, & T_{ij} \geq T_{ij}^c,
 \end{cases}
 \label{eq:de-Gennes}
\end{equation}
where $I_{ij}^c(0)$ is the junction’s critical current at $T = 0$; above $T_{ij}^c$ the segment is normal and the critical current vanishes. We control the degree of disorder in the arrays by considering a quenched normal distribution $\chi_{ij} \in \mathcal{N}(0, \sigma)$ where variables match the junctions’ labels in each array with zero mean and standard deviation $\sigma = 0.1$ as a generator for the other system’s observables. Generating a unique $\chi_{ij}$ distribution for each network allows us to model the sample-to-sample disorder level fluctuations observed in the experiment due to the experimental sample fabrication method. In particular, we define $I_{ij}^c(0) = I_0^c(1 + \chi_{ij})$ and $T_{ij}^c = T_c(1 + \chi_{ij})$, where the parameters $I_0^c = 58 \mu A$, $T_c = 2.9K$, and $R_n = 0.5 k\Omega$ were used in simulations for the main network. For the second network, used as the coupled network in Fig.~\ref{fig4}, we used $I_{0,2}^c = 48 \mu A$, $T_c = 2K$, and $R_n = 0.5 k\Omega$. All simulation measurements were performed on a 2D network of length $L=100$.

\subsection*{Thermal coupling}

To quantify the self-thermal coupling within the network, we evaluate the power dissipated by each resistor and its influence on the surrounding resistors. The power generated by segment $(i,j)$ is given by the standard electrical power relation:
\begin{equation}
 P_{ij} = \frac{V_{ij}^{2}}{R_{ij}},
 \label{eq:electric_power}
\end{equation}
where $V_{ij}$ and $R_{ij}$ denote the voltage across the resistor and its resistance, respectively.

To determine the thermal contribution of each resistor to all others, we use the discrete space version of the heat equation:
\begin{equation}
 \frac{\partial P}{\partial t} = D \nabla^{2} P,
 \label{eq:diffusion}
\end{equation}
where $D$ is the thermal diffusion constant.

The local power $Q_{ij}$ “experienced” by resistor $(i,j)$ is obtained by summing the thermally diffused contributions from all resistors in the network, using the two-dimensional Green’s function solution of the heat equation:
\begin{equation}
 Q_{ij}^{t}
 = \sum_{(k,l)} P_{ij,kl}^{t}
 = \sum_{(k,l)} \frac{P_{kl}^{t}}{4\pi D t}
 \exp\!\left(-\frac{|i-k|^{2} + |j-l|^{2}}{4Dt}\right),
 \label{eq:diffusion_sol}
\end{equation}
where $|i-k|$ and $|j-l|$ represent the two-dimensional distances between the emitting and receiving resistors. The lattice constant is set to unity. Periodic boundary conditions are imposed to eliminate edge-related effects. The discrete kernel on the periodic lattice is renormalized so that $\sum_{(k,l)} P_{ij,kl}^{t}/P_{kl}^{t} = 1$, ensuring conservation of the total dissipated heat. In the limit $Dt \to 0$ the kernel reduces to a Kronecker delta ($r=0$: heat remains on the emitting segment), which is the definition used for the $r=0$ case in Fig.~\ref{fig4}, where the expression $1/(4\pi Dt)$ is otherwise singular.

The product $Dt$ effectively characterizes the spatial scale over which heat diffuses and can thus be interpreted as a single diffusion parameter. Here $t$ denotes the diffusion time entering the kernel, which is fixed and distinct from the iteration index of the relaxation loop described below. When the analysis aims for a uniform heat distribution consistent with a mean-field approximation, the diffusion parameter $Dt$ is chosen to be large compared to the characteristic system size. The diffusion equation, solved using the spectral method, is efficiently evaluated with low numerical cost via Fast Fourier Transform (FFT), making it a computationally advantageous approach for modeling heat dispersion in our network system.

The thermal self-coupling is governed by the properties of the substrate, such that global thermal coupling determines the quasi-static state of the system. The updated global effective temperature at the $t$-th overheating cascade is therefore given iteratively by
\begin{equation}
 \mathbf{T}_{ij}^t = T_0 + \gamma Q_{ij}^t,
 \label{eq:global_thermal_coupling}
\end{equation}
where $T_0$ is the temperature of the heat bath and $\gamma$ is the heat retention coefficient of the network (the parameter used in simulations). Consequently, the heat retention coefficient is inversely proportional to the thermal conductivity of the network substrate.
In the simulations, we used $\gamma = 10^8~\mathrm{W^{-1}\,K}$ as the heat retention coefficient of the self-coupled network and $\gamma =4 \times 10^{3}~ \mathrm{W^{-1}\,K}$ as the heat retention coefficient within and between the coupled network in Fig.~\ref{fig4}.

\subsection*{Thermally coupled Kirchhoff equations}

We developed a model of a thermally self-coupled RSJJ network in which thermal interactions arise from the heat dissipated by individual junctions. Numerical solutions for the order parameter (here, the global sheet resistance $R$) are obtained recursively by allowing the layer to self-interact through its isolated behavior adaptively \cite{ponta2009resistive}. In our model of a self-coupled thermal RSJJ network, this is accomplished by solving the Kirchhoff equations of the array under the adaptive interplay between ~\eqref{eq:JJ_characteristics}, \eqref{eq:de-Gennes}, and \eqref{eq:diffusion_sol}. We consider a two-dimensional lattice of linear size $L$, with its left and right boundaries connected to an external super-node (source) for current injection and to ground, respectively. Each junction follows a Josephson $I$–$V$ characteristic with $R_{ij}$ defined as in~\eqref{eq:JJ_characteristics}, where we take $R_{\epsilon} = 10^{-5}~\Omega$ for the SC state resistance and $R_n = 0.5~\mathrm{k}\Omega$ for the N state resistance. The algorithm is initialized by assigning resistances based on the phase state: when starting from the mutual SC phase, we set $R_{ij,0} = R_{\epsilon}$, whilst for an initial mutual N phase, we set $R_{ij,0} = R_n$.

The algorithm evolves iteratively as follows:\\
\noindent
(i) In the $t$-th stage ($t \geq 1$) of the overheating cascade, the symmetric conductance matrix $\mathbf{G}_t$ is constructed from the junction resistances $R_{ij,t-1}$, with entries
$$
G_{ij}=\begin{cases}
 0, & \text{if $(i,j) \notin E$},\\
 -1/R_{ij}, & \text{if $(i,j) \in E$},\\
 \sum_{k \in \partial i} 1 / R_{ik}, & \text{if $i = j$},
\end{cases}
$$
where $E$ is the set of edges in the array and $\partial i$ is the set of nearest neighbors of node $i$;\\

\noindent
(ii) The potential vector $\mathbf{V}_{t}$ is obtained by numerically solving the Kirchhoff matrix equation
\begin{equation}
 \mathbf{G}_t \cdot \mathbf{V}_{t} = \mathbf{I}_{\mathrm{inj}},
\end{equation}
where ($\cdot$) denotes matrix multiplication and $\mathbf{I}_{\mathrm{inj}}$ is the vector of injected nodal currents, whose entries are zero except for the first component (the super-node) which equals the driving current $I_T$;\\

\noindent
(iii) The heat emitted by each junction is computed through its electrical power, using ~\eqref{eq:electric_power}, with the resistance $R_{ij,t-1}$ and the potential drop $V_{ij,t}$ extracted from $\mathbf{V}_t$;\\

\noindent
(iv) Heat is subsequently thermally diffused across the network. Using the spectral method to solve~\eqref{eq:diffusion}, the contributions are summed via~\eqref{eq:diffusion_sol}, and the effective temperature $\mathbf{T}_t$ is computed using~\eqref{eq:global_thermal_coupling};\\

\noindent
(v) The critical currents $I_{ij}^c(T)$ are updated using~\eqref{eq:de-Gennes}, and the resistive state $R_{ij,t}$ of each junction is determined according to~\eqref{eq:JJ_characteristics};\\

\noindent
(vi) The global resistance of the network is computed as $R_t = \mathbf{V}_t(N) / I_T$.\\

\par
Steps (i)–(vi) are repeated recursively to generate a sequence of potential vectors $\{\mathbf{V}_{1}, \dots, \mathbf{V}_{t}, \dots\}$. Convergence is established once the error,
\begin{equation*}
 \delta V = \max_{ij} \left| 1 - \frac{\mathbf{V}_{t}}{\mathbf{V}_{t+1}} \right|,
\end{equation*}
falls below a numerical threshold $\epsilon_{\min}$, where the ratio is taken component-wise and the maximum is over all nodes with $\mathbf{V}_{t+1}\neq 0$. In the simulations, we used $\epsilon_{\min} = 10^{-5}$ for cooling and $\epsilon_{\min} = 10^{-8}$ for heating.

\clearpage
\onecolumngrid
\raggedbottom

\begin{center}
{\Large \textbf{Supplementary Information}}
\end{center}

\renewcommand{\thefigure}{S\arabic{figure}}
\setcounter{figure}{0}
\renewcommand{\theequation}{S\arabic{equation}}
\setcounter{equation}{0}
\renewcommand{\thetable}{S\arabic{table}}
\setcounter{table}{0}

This Supplementary Information documents additional analyses supporting the
main text, and is organized into four Supplementary Notes.
Supplementary Note~1 establishes that the phase transition reported in the main
text is a well-defined, reproducible transition rather than an artifact of a
single realization or of a particular measurement convention: it introduces
four operationally distinct definitions of the critical temperature $T_c$,
shows that they yield consistent estimates of the same transition temperature,
and clarifies that the small-scale steps seen in individual resistance curves
are a single-realization, finite-size feature, smoothed by ensemble averaging
and distinct from the macroscopic transition discontinuity, when such a
discontinuity is present.
Supplementary Note~2 presents the ensemble-averaged $T_c(r)$ of the main-text
Fig.~2b under all four definitions.
Supplementary Note~3 presents the range-induced stabilization of the main-text
Fig.~4 using the threshold definition, which is the most direct probe of the
effect.
Supplementary Note~4 provides extended results and illustrations for the
two-layer interdependent (ISN) and single-layer self-dependent (SDSN)
configurations.
Supplementary Notes~1 and~2 use the single-layer SDSN; Supplementary Note~3
uses the two-layer heating protocol (see the note at its start). Unless stated
otherwise, all results in Supplementary Notes~1--3 use $100$ realizations per
parameter set.

\section*{Supplementary Note 1: Operational definition of the critical temperature}
\label{sec:Tc_definition}
A recurring difficulty in disordered, avalanche-driven transitions is that the
location of $T_c$ can appear to depend on the criterion used to identify it. To
demonstrate that this is not the case here, we define $T_c$ in four
operationally distinct ways, each probing a different physical signature of the
collapse:

\begin{enumerate}
  \item \textbf{Maximal derivative} (\texttt{deriv}): the temperature at which
        $|\mathrm{d}R/\mathrm{d}T|$ is maximal, i.e.\ the steepest point of the
        resistance change.
  \item \textbf{Resistance threshold} (\texttt{thresh}): the temperature at
        which the normalized resistance crosses a fixed small threshold
        ($R=0.1$ of the normal-state value). During cooling this is the
        temperature at which $R$ first falls below $0.1$; during heating it is
        the temperature at which $R$ first rises above $0.1$.
  \item \textbf{Spanning cluster} (\texttt{cluster}): the temperature at which a
        superconducting cluster first spans the system during cooling (or, under
        heating, at which the spanning superconducting cluster is lost). The
        cluster criterion is used here for the cooling protocol of Supplementary Notes~1--2.
  \item \textbf{Relaxation peak} (\texttt{iter}): the temperature at which
        the number of iterations required for the self-consistent Kirchhoff
        solution to converge is maximal, i.e.\ where the convergence time is
        longest near the transition.
\end{enumerate}

These criteria are operationally distinct: two are based on the macroscopic
transport curve, one on the geometry of the superconducting phase, and one on
the convergence time of the self-consistent solution. Agreement among them is
therefore a strong, non-trivial indication that they all detect the same
underlying transition.

\begin{figure}[!htb]
  \centering
  \includegraphics[width=0.9\linewidth]{ 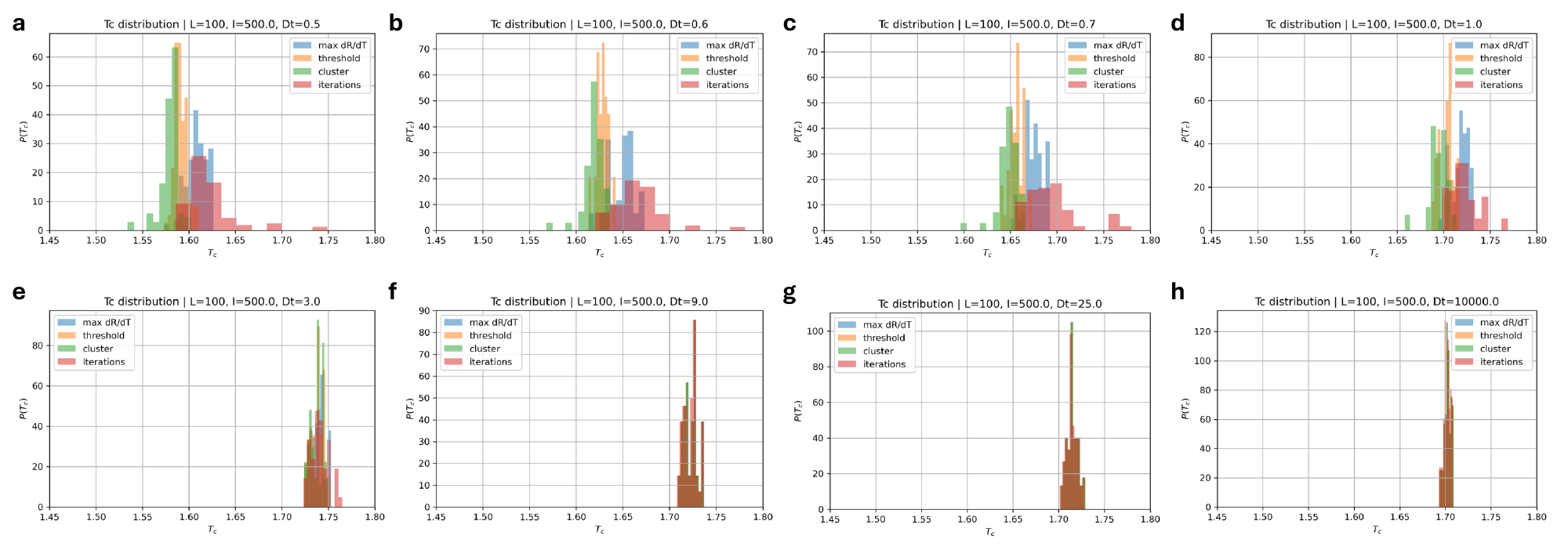}
  \caption[Distributions of $T_c$ under four definitions]{%
  \textbf{Distribution of $T_c$ over realizations, obtained from four
  operationally distinct definitions.}
  Histograms of $T_c$ across $100$ realizations for $L=100$, $I=500\,\mu$A,
  using the four criteria of Supplementary Note~1
  (\texttt{max dR/dT}, \texttt{threshold}, \texttt{cluster}, \texttt{iterations}).
  \textbf{a-d.} Short-to-intermediate ranges ($\mathrm{Dt}=0.5,0.6,0.7,1.0$; recall
  $r=\sqrt{\mathrm{Dt}}$). The four distributions overlap and identify a common $T_c$
  within their widths; the \texttt{cluster} and \texttt{iterations} criteria
  are somewhat noisier and slightly broader at the smallest $\mathrm{Dt}$, where the
  transition is continuous and the collapse is least sharply localized.
  \textbf{e-h.} Larger ranges ($\mathrm{Dt}=3,9,10^4$). As the range increases the four
  definitions converge tightly onto a single value, consistent with the
  progressively sharper (nucleation- and mixed-order) collapse.
  }
  \label{fig:Tc_histograms}
\end{figure}

Figure~\ref{fig:Tc_histograms} shows that the four definitions yield consistent
estimates of $T_c$. At small $\mathrm{Dt}$ (continuous transition) the
distributions are broader and the \texttt{cluster}/\texttt{iterations} criteria
are noisier, but all four agree within their widths; at large $\mathrm{Dt}$ the
distributions collapse onto a single sharp value. We therefore adopt the
\textbf{threshold} definition as the primary criterion in the main text: it is
numerically the most stable across the full range of $\mathrm{Dt}$ and it is the most
direct probe of the quantity that the stabilization effect actually modifies,
namely the resistive threshold of the network.

\begin{figure}[!htb]
  \centering
  \includegraphics[width=0.8\linewidth]{ 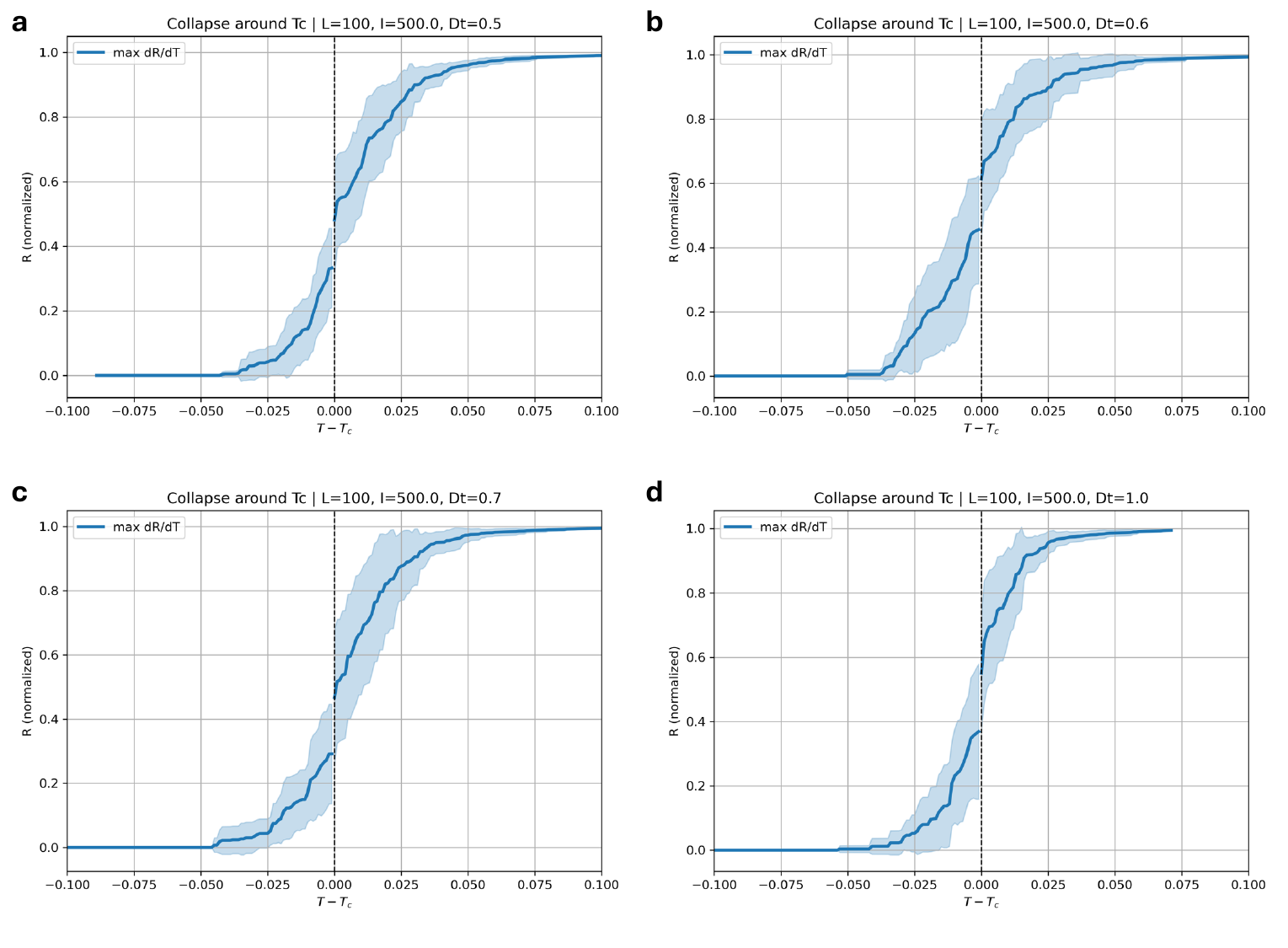}
  \caption[$T_c$-aligned resistance curves]{%
  \textbf{Realization-resolved, $T_c$-aligned resistance curves.}
  Normalized resistance as a function of $T-T_c$ for $L=100$, $I=500\,\mu$A, at
  several values of $\mathrm{Dt}$, where $T_c$ is determined per realization
  (\texttt{max dR/dT} definition). The solid line is the ensemble mean and the
  shaded band its realization-to-realization spread. Aligning each realization
  to its own $T_c$ demonstrates that the transition is sharply defined relative
  to a well-localized critical temperature.
  }
  \label{fig:collapse}
\end{figure}

\subsection*{Small-scale steps: a single-realization, finite-size feature}
\label{sec:jumps_artifact}
Individual resistance curves display small step-like features superimposed on
the main change. These steps are \emph{not} the macroscopic transition
discontinuity and should not be interpreted as such. They are a property of a
single, finite realization: they reflect the discrete reconfiguration events of
a finite network, and they are smoothed out under ensemble averaging and
diminish with increasing system size $L$. The averaged curves shown here (see
Supplementary Note~2) are therefore free of these steps.

A methodological caution is worth stating explicitly, because it bears directly
on how averaged curves should be read. There are two distinct averaging
procedures, with opposite pitfalls:
\begin{itemize}
  \item Averaging many realizations \emph{at fixed absolute temperature} $T$
        removes the single-realization steps and yields the smooth macroscopic
        transition. This is the procedure used for the averaged curves.
  \item Averaging many realizations \emph{after aligning each to its own}
        $T_c$ (i.e.\ as a function of $T-T_c$) is useful for exposing the
        intrinsic sharpness of the collapse (Fig.~\ref{fig:collapse}), but it
        must be used with care: for a genuinely continuous (second-order)
        transition this alignment can \emph{induce} an apparent jump at
        $T-T_c=0$ that is an artifact of the alignment rather than a physical
        discontinuity. We therefore restrict $T_c$-aligned averaging to
        diagnostic use and never infer the order of the transition from it.
\end{itemize}
This distinction is what allows us to separate the physical transition
discontinuity (present for intermediate and long ranges, where the transition is
abrupt) from both the finite-size steps and the alignment artifact.

\section*{Supplementary Note 2: Ensemble-averaged $T_c(r)$ (main-text Fig.~2b)}
\label{sec:Tc_r}
\begin{figure}[!htb]
  \centering
  \includegraphics[width=1\linewidth]{ 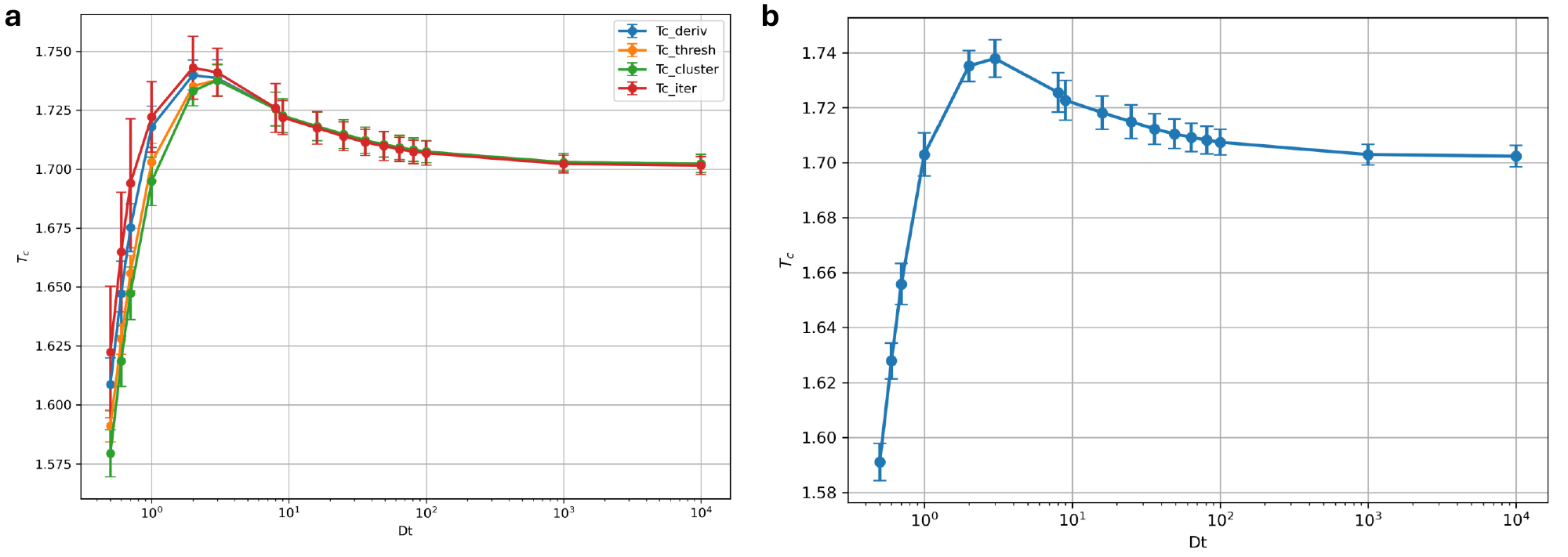}
  \caption[Ensemble-averaged $T_c(r)$]{%
  \textbf{Ensemble-averaged critical temperature as a function of interaction
  range.}
  $T_c$ versus $r=\sqrt{\mathrm{Dt}}$ for $L=100$, $I=500\,\mu$A, averaged over $100$
  realizations; error bars denote the realization-to-realization standard
  deviation.
  \textbf{a.} All four definitions. $T_c$ rises steeply at small $r$, reaches a
  maximum near $r\simeq\sqrt{3}$, and then decreases and saturates at large $r$.
  The four definitions differ slightly in the small-$r$ (continuous) regime but
  coincide at and beyond the maximum, where the collapse is sharp.
  \textbf{b.} The threshold definition alone, as used in the main-text Fig.~2b.
  The non-monotonic dependence of $T_c$ on range, an increase followed by a
  decrease toward a plateau, parallels the behavior of geometric
  interdependent networks. The distinguishing physical feature is not this
  non-monotonicity itself, but the range--strength locking (Supplementary Note~3), which is
  absent from geometric models and which, in the heating protocol, allows
  increasing range to stabilize the superconducting phase, as demonstrated by
  comparing the diffusive curves with the $r=0$ baseline
  (Fig.~\ref{fig:heating_curves}).
  }
  \label{fig:Tc_vs_r}
\end{figure}

Figure~\ref{fig:Tc_vs_r} presents the critical temperature as a function of the
interaction range $r=\sqrt{\mathrm{Dt}}$, averaged over $100$ realizations, using all
four definitions of Supplementary Note~1. The main-text Fig.~2b shows
the threshold definition alone (Fig.~\ref{fig:Tc_vs_r}b); the full four-way
comparison is given here.

\begin{figure}[!htb]
  \centering
  \includegraphics[width=0.7\linewidth]{ 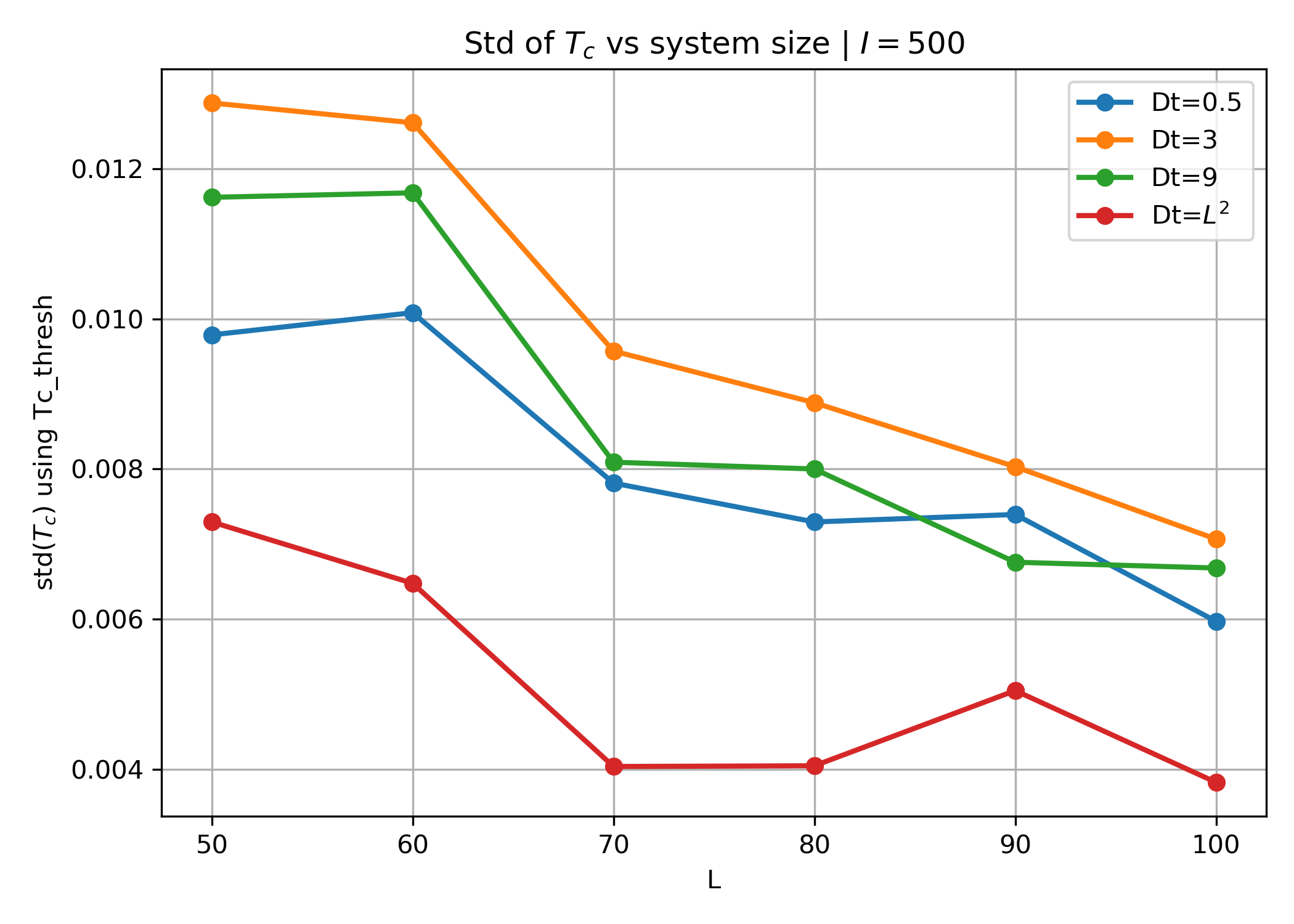}
  \caption[Standard deviation of $T_c$ versus system size]{%
  \textbf{Sample-to-sample fluctuations of $T_c$ show an overall decrease with
  system size.}
  Standard deviation of $T_c$ across realizations (\texttt{threshold} definition)
  as a function of $L$, for several thermal ranges $\mathrm{Dt}$, at fixed total
  current $I=500\,\mu$A. The overall trend is decreasing, consistent with
  convergence toward a sharp transition in the thermodynamic limit, with
  substantial realization-to-realization and finite-sample fluctuations. The
  case $\mathrm{Dt}=L^2$ (red) is a size-dependent, near-uniform-heating choice
  (not a fixed $\mathrm{Dt}$) and shows the smallest fluctuations.
  }
  \label{fig:std_vs_L}
\end{figure}

The realization-to-realization scatter of $T_c$ shows an overall decrease with
increasing system size. Within the accessible sizes, this is consistent with a
transition that sharpens toward a well-defined critical point in the
thermodynamic limit.
Figure~\ref{fig:std_vs_L} shows the standard deviation of $T_c$ (from the
\texttt{threshold} definition) as a function of $L$ for several values of
$\mathrm{Dt}$ ($\mathrm{Dt}=0.5,3,9$ and the size-dependent $\mathrm{Dt}=L^2$):
the overall trend is decreasing, with substantial
realization-to-realization and finite-sample fluctuations superimposed. The case
$\mathrm{Dt}=L^2$ is not a fixed value of $\mathrm{Dt}$ but a size-dependent
choice that keeps the diffusion range comparable to the system size (near-uniform
heating) for every $L$, and it gives the smallest and most monotonic scatter. All
size comparisons in Fig.~\ref{fig:std_vs_L} were performed at fixed total applied
current $I=500\,\mu$A, so they test convergence under a fixed-drive
protocol rather than a strict current-density-preserving finite-size scaling.

\section*{Supplementary Note 3: Range-induced stabilization (main-text Fig.~4)}
\label{sec:stabilization}
Unlike Supplementary Notes~1--2, which use the single-layer SDSN, this note uses the
two-layer heating protocol (two thermally coupled layers) unless explicitly
stated otherwise.

The main-text Fig.~4 reports the central conceptual result of this work: in a
physically coupled network, extending the dependency range can \emph{raise} the
critical temperature, i.e.\ dependency can \emph{stabilize} the superconducting
phase. We stress that this is presented as an \textbf{existence result}, not as
a claim of generic behavior: naively one expects long-range dependency to
destabilize the system, as it does in geometric interdependent networks, yet
here a single physically reasonable setup suffices to establish that the
opposite outcome is possible. (Each $T_c(r)$ point below is itself averaged over
many disorder realizations; ``existence'' refers to the physical setup and
parameter regime, not to a single random realization.)

The mechanism is a property of physical, diffusion-mediated dependency and has
no counterpart in geometric models. In a geometric network the coupling range
and coupling strength are independent parameters. In a diffusive physical
network they are inseparably linked: for a fixed heat input and a
normalized diffusion kernel, spreading the heat over a larger range dilutes its
local amplitude at every point, so that increasing the range \emph{necessarily}
weakens the local coupling. While $r=\sqrt{\mathrm{Dt}}$ sets only the width of
the kernel, the peak amplitude of the normalized kernel scales as
$K(0)=1/(4\pi \mathrm{Dt})=1/(4\pi r^2)$, so the local coupling strength falls off
as $1/r^2$ as the range grows. This range--strength locking, acting together with
the current redistribution imposed by Kirchhoff's laws, is what reverses the usual
role of dependency.

\begin{figure}[!htb]
  \centering
  \includegraphics[width=0.7\linewidth]{ 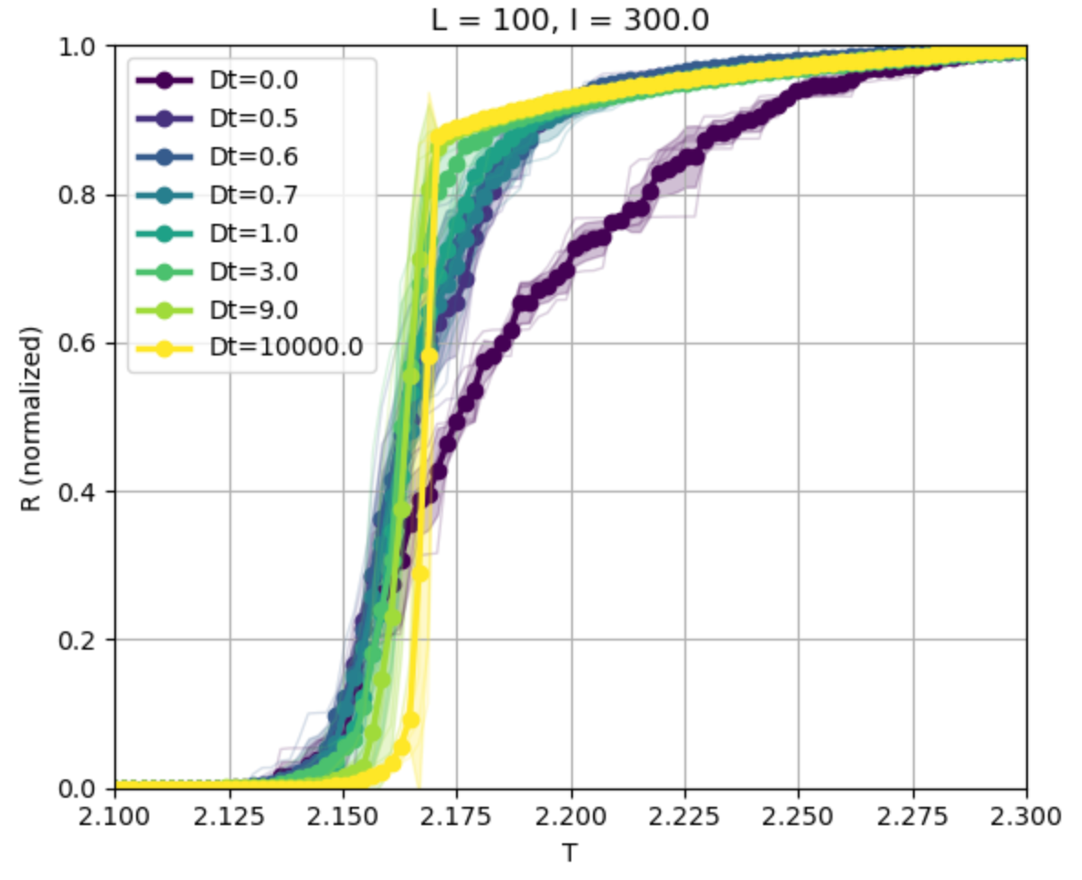}
  \caption[Two-layer heating resistance curves]{%
  \textbf{Resistance curves for two-layer heating.}
  Normalized resistance versus temperature during heating for two thermally
  coupled layers, $L=100$, $I=300\,\mu$A, for several thermal ranges $\mathrm{Dt}$
  (with $r=\sqrt{\mathrm{Dt}}$; $\mathrm{Dt}=0$ denotes the no-diffusion baseline). Light lines
  are individual realizations; heavy lines with markers are ensemble averages
  over $100$ realizations.
  }
  \label{fig:heating_curves}
\end{figure}

\begin{figure}[!htb]
  \centering
  \includegraphics[width=0.6\linewidth]{ 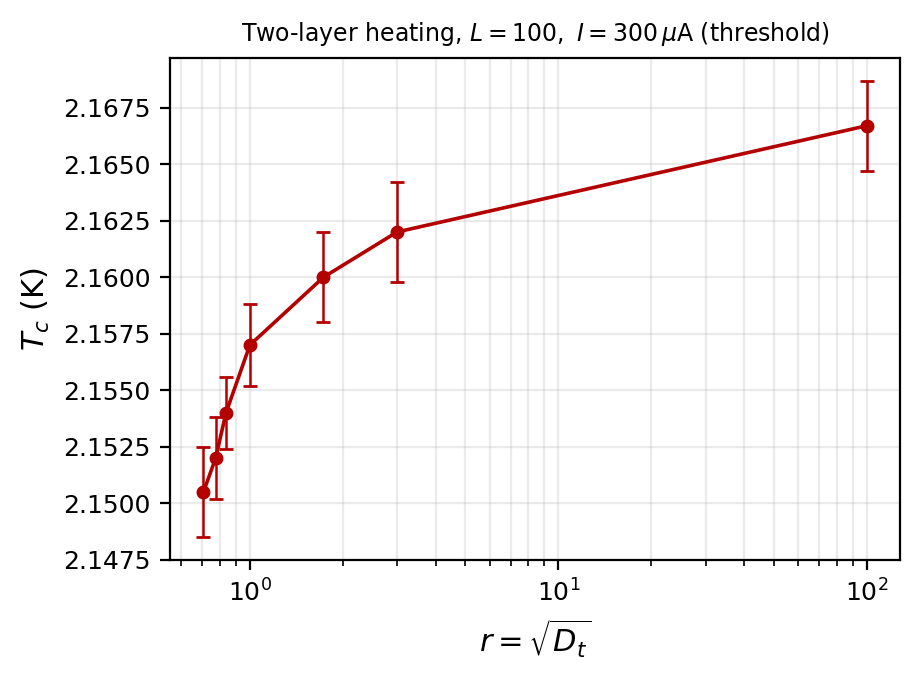}
  \caption[Range-induced stabilization, threshold definition]{%
  \textbf{Critical temperature increases with interaction range under heating
  (threshold definition).}
  $T_c$ versus $r=\sqrt{\mathrm{Dt}}$ for the two-layer heating protocol, $L=100$,
  $I=300\,\mu$A, threshold definition, averaged over realizations; error bars
  denote the realization-to-realization standard deviation. $T_c$ increases
  monotonically with range and saturates at large $r$: extending the dependency
  range stabilizes the superconducting phase \emph{within the diffusive regime}.
  The horizontal axis is logarithmic, so the $\mathrm{Dt}=0$ (no-diffusion,
  $r=0$) baseline of Fig.~\ref{fig:heating_curves} cannot be shown here; the
  plotted range therefore begins at $\mathrm{Dt}=0.5$. The $\mathrm{Dt}=0$
  baseline is shown separately in the resistance curves of
  Fig.~\ref{fig:heating_curves} and should be read from there. This is the
  quantity shown in the main-text Fig.~4, and it constitutes an existence
  demonstration of dependency-induced stabilization with increasing range rather
  than a claim of universal behavior.
  }
  \label{fig:Fig4_Tc}
\end{figure}

Figure~\ref{fig:heating_curves} shows the two-layer heating resistance curves
that underlie Fig.~4, and Fig.~\ref{fig:Fig4_Tc} shows the corresponding
$T_c(r)$ using the threshold definition. We use the threshold definition here
for the same reason as in the main text: it is the most direct probe of the
resistive stabilization and yields the cleanest signal. The effect is present
but weaker under the \texttt{deriv} and \texttt{iterations} definitions, where
part of the rise lies within the error bars; this is expected, since those
criteria are less directly tied to the resistive threshold that the
stabilization modifies.

\section*{Supplementary Note 4: Additional results for the two-layer (ISN) and single-layer (SDSN) systems}
The following figures provide extended results and illustrations that support
the analysis of the main text. The canonical transition regimes and the principal
cascade-dynamical features appear in both the single-layer self-dependent
superconducting network (SDSN) used in the main text and in the two-layer
interdependent superconducting network (ISN).

\subsection*{Diffusion kernel and range--strength locking}
\begin{figure*}[htbp]
    \centering
    \includegraphics[width=1\linewidth]{ 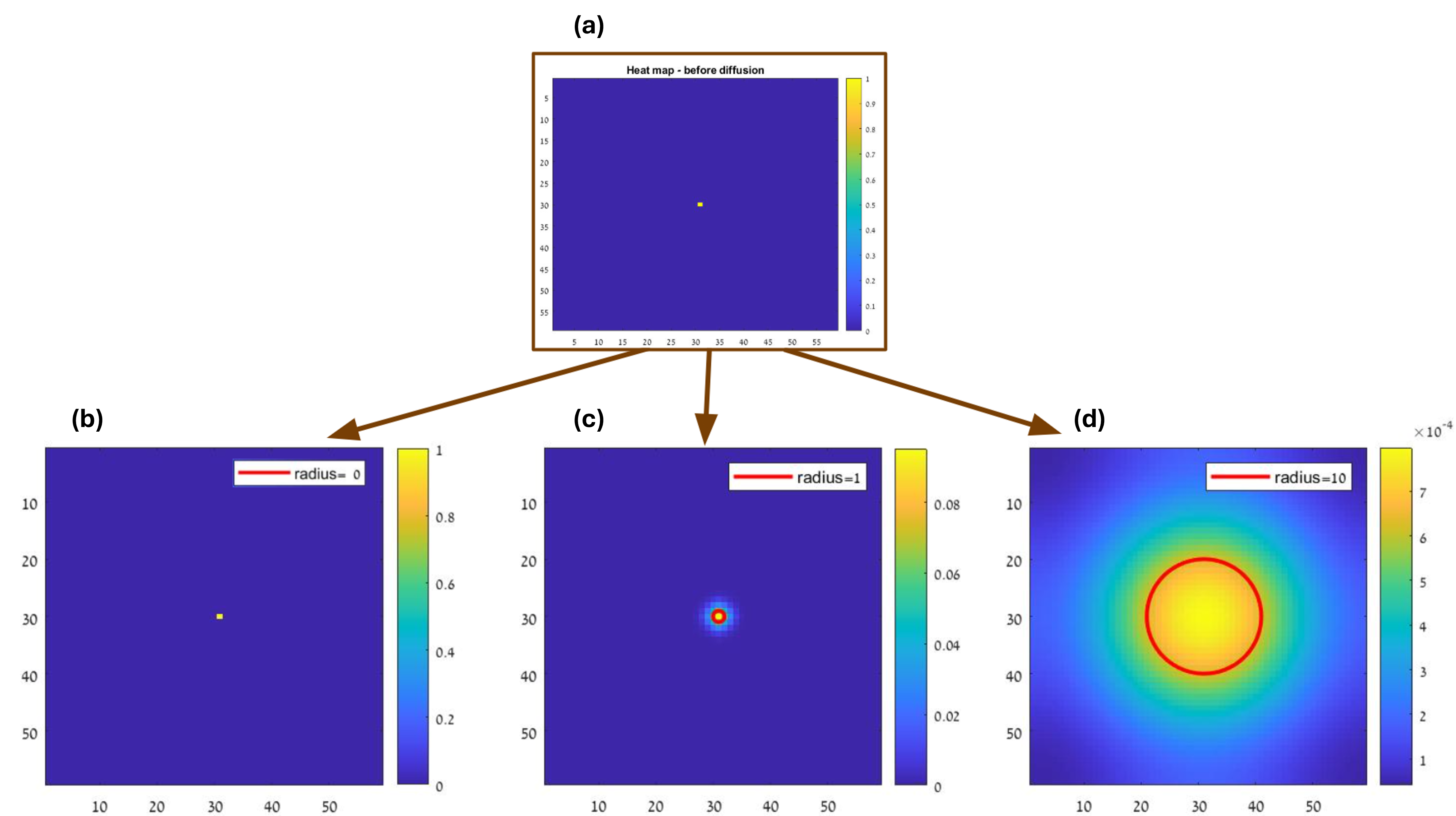}
    \caption[Demonstration of heat diffusion]{
\textbf{Illustration of heat diffusion emitted by a single resistive element up to a finite interaction radius $r$.}
(\textbf{a}) Initial heat map before diffusion: all resistors in the network are in the superconducting phase, except for a single resistive junction located at the center, which generates Joule heat.
(\textbf{b}) Heat map after diffusion for $r=0$. The heat remains strictly localized at the emitting resistor, indicating that no diffusion occurs and the generated heat affects only the resistor itself.
(\textbf{c}) Heat map after diffusion for $r=1$. The heat spreads weakly to the nearest neighbors, remaining largely localized. The reduced intensity scale (shown on the color bar) reflects the conservation of total heat, which is redistributed over a small region, leading to a lower temperature at the emitting resistor.
(\textbf{d}) Heat map after diffusion for $r=10$. The heat spreads over a much larger portion of the system, approaching an almost uniform distribution across the network. The significantly narrower intensity range indicates a near mean-field--like regime, where the heat is distributed nearly homogeneously throughout the system.
}
    \label{fig:SI_heatdiff}
\end{figure*}

Figure~\ref{fig:SI_heatdiff} illustrates the heat-diffusion kernel and the way the
interaction range and coupling strength are locked together
(see Supplementary Note~3).

\subsection*{Two-layer geometry and collapse dynamics}
\begin{figure*}[htbp]
    \centering
    \includegraphics[width=1\linewidth]{ 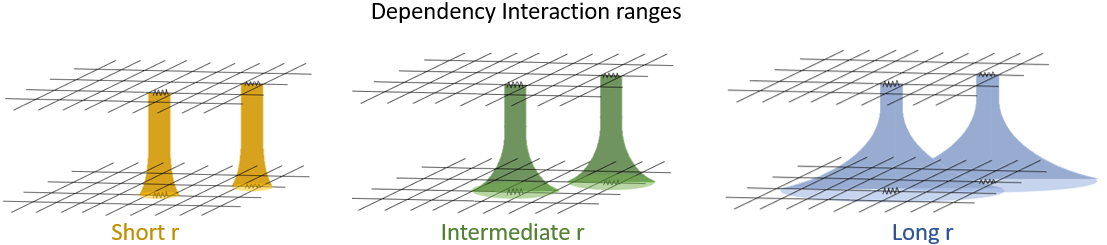}
    \caption[Illustration of different scales of heat diffusion in ISN]{
\textbf{Schematic illustration of interdependent superconducting networks with diffusive heat interactions at different length scales}.
Two spatially embedded networks of amorphous superconducting segments are arranged in parallel and separated by an electrically insulating but thermally conducting medium. The layers interact through Joule heating with a thermal interaction range $r$.
Each segment $(i,j)$ of the two-dimensional lattices represents a link modeled as a resistor-shunted Josephson junction (RSJJ), characterized by a local critical current $I^c_{i,j}(T)$.
As the temperature varies, individual segments may undergo transitions between the superconducting and normal states.
Zigzag symbols denote resistive (normal) segments that dissipate Joule heat, which subsequently diffuses through the thermal medium with diffusion constant $D$.
The colored regions illustrate the effective thermal interaction radius $r=\sqrt{Dt}$, representing the typical spatial range over which heat emitted by a resistive segment affects corresponding locations in the parallel network, where $t$ is set by the characteristic thermal response time of the system.}
    \label{fig:SI_Ilustration_2Net}
\end{figure*}

\begin{figure*}[htbp]
    \centering
    \includegraphics[width=1\linewidth]{ 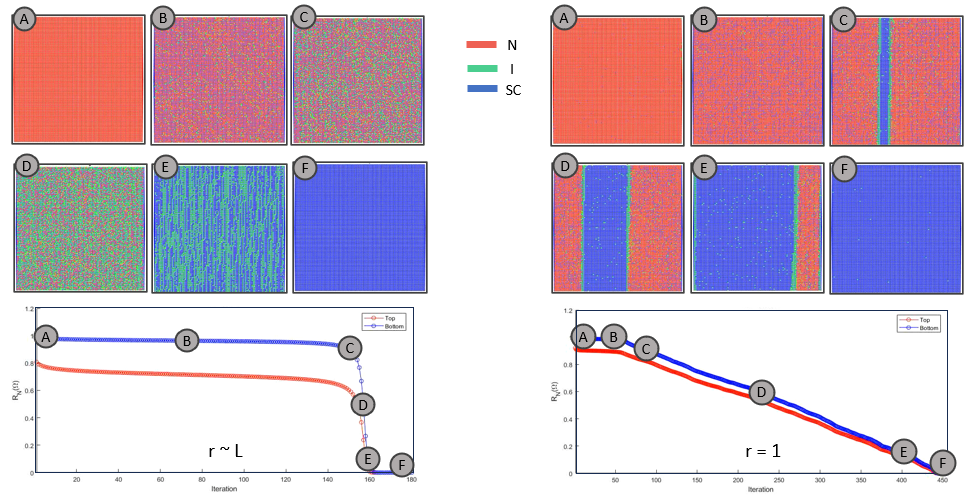}
      \caption[Nucleation vs. random cascade in ISN]{
\textbf{Spontaneous cascading dynamics of the collapse at $T_c$.}
The total resistance of one network in a system of two interdependent superconducting networks is shown as a function of iteration number during cooling from the conducting state at $T = 3\,$K to a temperature just after criticality, under a constant applied current $I=300\,\mu\mathrm{A}$.
Left panels correspond to the long-range interaction limit $r= L = 100$ ($T_c=1.321\,$K), associated with a mixed-order transition, while right panels correspond to the short-range case $r=1$ ($T_c=1.4\,$K), exhibiting first-order--like behavior.
The resistance curves and snapshots (A--F) correspond to a single disorder realization, chosen to display the spatial structure of the collapse; the snapshots show the two-dimensional resistor network at representative stages, as marked along the curves.
Colors denote the local state of each segment: normal (N, red), superconducting (SC, blue), and intermediate (I, green) (see Methods).
Further discussion of the collapse mechanisms is provided in the main text.
}
    \label{fig:SI_screenshots_2Net}
\end{figure*}

\begin{figure*}[htbp]
    \centering
    \includegraphics[width=1\linewidth]{ 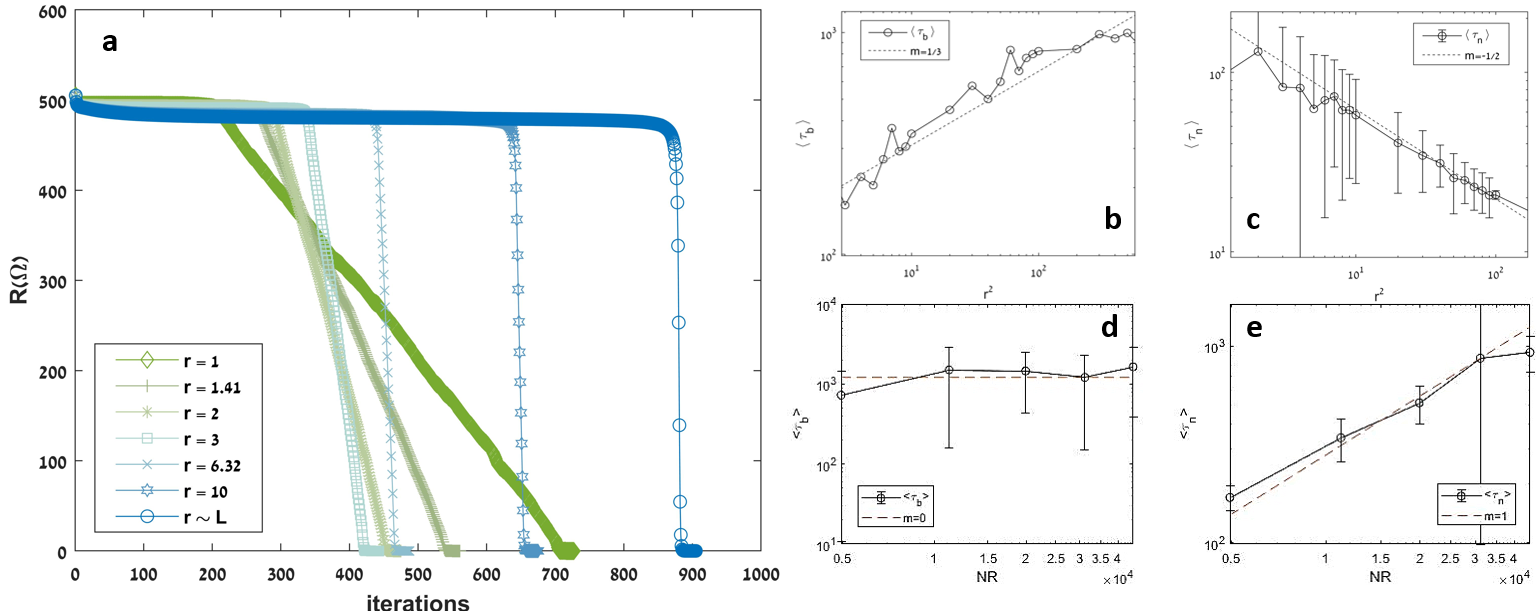}
    \caption[Dynamics analysis in ISN]{
\textbf{Dynamics of the resistance collapse at criticality for different thermal interaction ranges (ISN).}
\textbf{(a)} Network resistance $R(\Omega)$ as a function of iteration number during cooling from the conducting state at $T=3\,\mathrm{K}$ to the critical temperature of each $r$, under a constant applied current $I=300\,\mu\mathrm{A}$.
Different curves correspond to different thermal interaction ranges $r$, spanning the nucleation regime and the long-range cascading regime.
The collapse proceeds via a plateau associated with a critical branching process, followed by a linear decrease of the resistance corresponding to nucleation-driven propagation.
Data in panels (b--e) are averaged over 16 realizations.
\textbf{(b)} Mean duration of the branching (plateau) regime at criticality, as a function of the squared thermal interaction range $r^2$. The duration of the plateau increases like $r^{2/3}$.
\textbf{(c)} Mean duration of the nucleation regime at criticality, as a function of $r^2$. The duration of the nucleation period decreases with $r^{-1}$.
\textbf{(d)} For $r=1$, the mean duration of the branching (plateau) regime at criticality as a function of the number of segments, $NR = 2\times L\times L$. The plateau duration is nearly independent of system size, consistent with a microscopic timescale associated with the initiation of nucleation.
\textbf{(e)} For $r=1$, the mean duration of the nucleation regime at criticality as a function of $NR$. The nucleation duration increases linearly with $NR$ (power-law exponent $m=1$).
}
    \label{fig:SI_dynamics_2Net}
\end{figure*}

Figure~\ref{fig:SI_Ilustration_2Net} presents the two-layer (ISN) geometry,
Figure~\ref{fig:SI_screenshots_2Net} the spatial structure of the collapse at
$T_c$ in this configuration, and Figure~\ref{fig:SI_dynamics_2Net} the
corresponding collapse dynamics and their scaling with $r$ and with system size.

\subsection*{Extended single-layer results}
\begin{figure*}[htbp]
    \centering
    \includegraphics[width=1\linewidth]{ 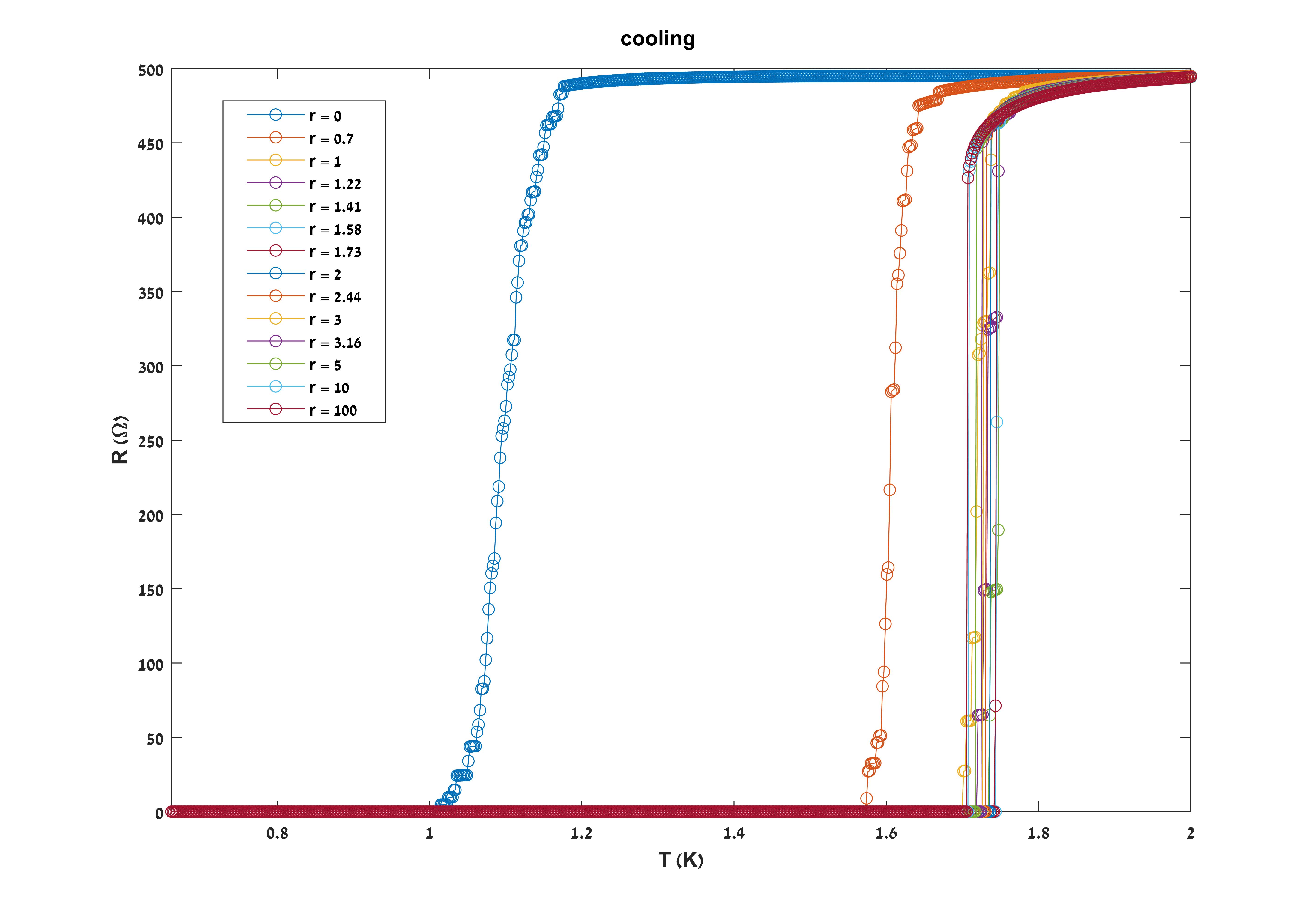}
    \caption[Cooling systems with different $r$ in SDSN]{\textbf{Resistance as a function of temperature during cooling (SDSN).} Resistance $R(\Omega)$ as a function of temperature $T(\mathrm{K})$ during cooling for a system of size $L = 100$, shown for different thermal interaction ranges $r$. These data represent the full results for a single self-dependent network (a single disorder realization). For small $r$, the resistance decreases gradually over a broad temperature interval, indicating a continuous transition. As $r$ increases, the transition becomes progressively sharper, eventually developing an abrupt jump characteristic of a first-order--like transition, while the critical temperature increases. For sufficiently large $r$, the critical temperature decreases again and converges to a limiting value, and the abrupt collapse turns into a mixed-order transition.
}
    \label{fig:SI_cooling_1Net}
\end{figure*}

\begin{figure*}[htbp]
    \centering
    \includegraphics[width=1\linewidth]{ 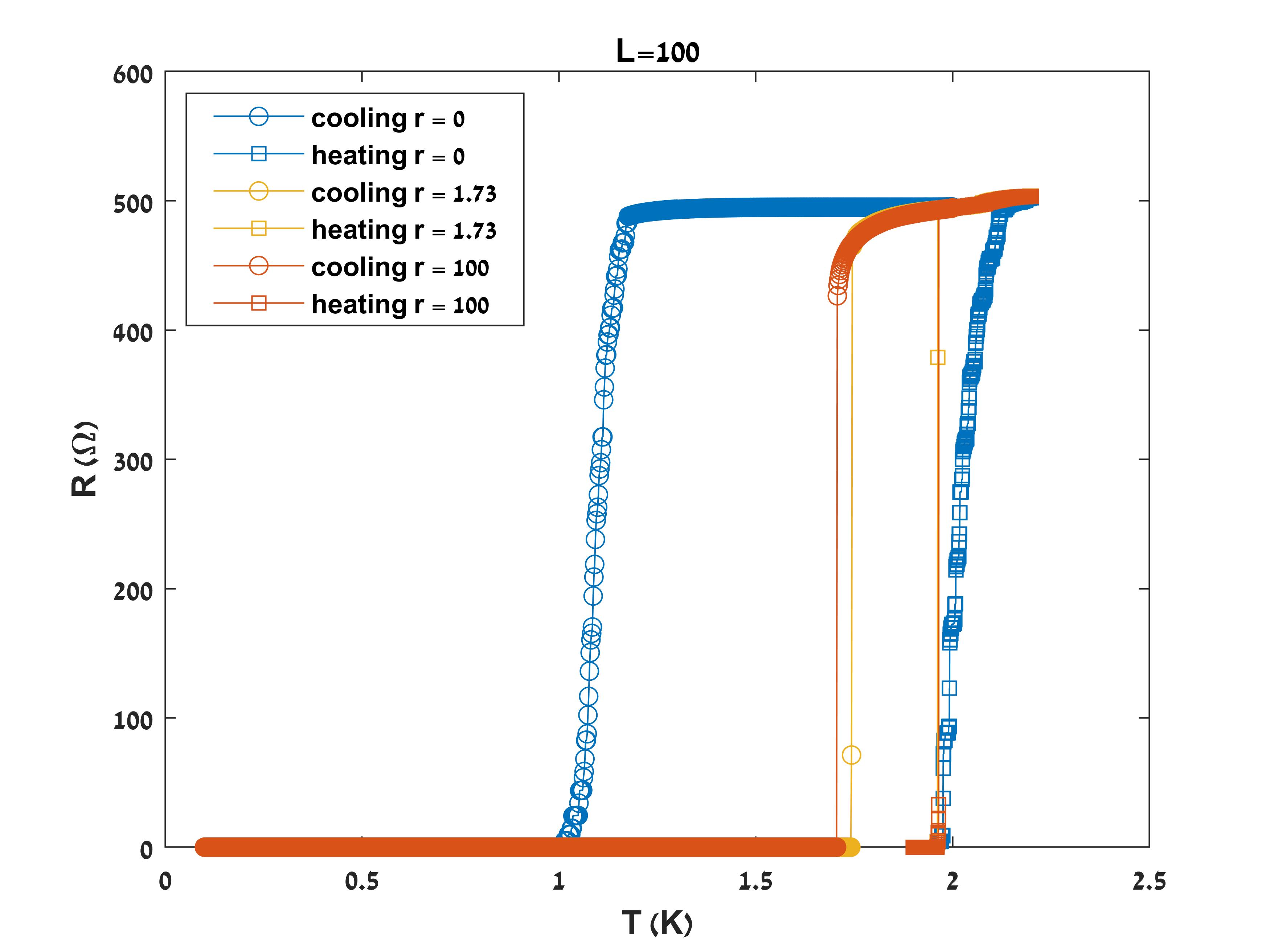}
    \caption[Hysteresis in systems with different $r$ in SDSN]{\textbf{Hysteresis in systems with different $r$ in SDSN.} The figure presents the resistance $R(\Omega)$ as a function of temperature $T(\mathrm{K})$ during heating and cooling for a system of size $L = 100$, shown for different thermal interaction ranges $r$ (short, medium and long interaction, single self-dependent network; single disorder realization). In addition to the expected hysteresis opening as a consequence of the abrupt phase transition observed for intermediate and long interaction ranges, a hysteresis (a shift between cooling and heating curve) also appears at short interaction ranges that give rise to a second-order transition. Although hysteresis is not commonly associated with second-order transitions, in the present system it is well understood and follows from the dependence of the critical behavior on the initial state of the system. When the system starts from the normal state (cooling), it dissipates heat, which affects the critical currents of the segments and forces them to undergo the phase transition at a lower temperature. In contrast, during heating, when the system starts from the superconducting state and does not dissipate power, the critical currents remain maximal, allowing the phase transition to occur at a higher temperature. A similar coexistence of a continuous transition with hysteresis has been reported in other systems in which spatial heterogeneity competes with nucleation dynamics, such as the finite-dimensional random-field Ising model, where hysteresis arises from local metastability and interface pinning rather than from a macroscopic avalanche.
}
    \label{fig:SI_hysteresis_1Net}
\end{figure*}

\begin{figure*}[htbp]
    \centering
    \includegraphics[width=1\linewidth]{ 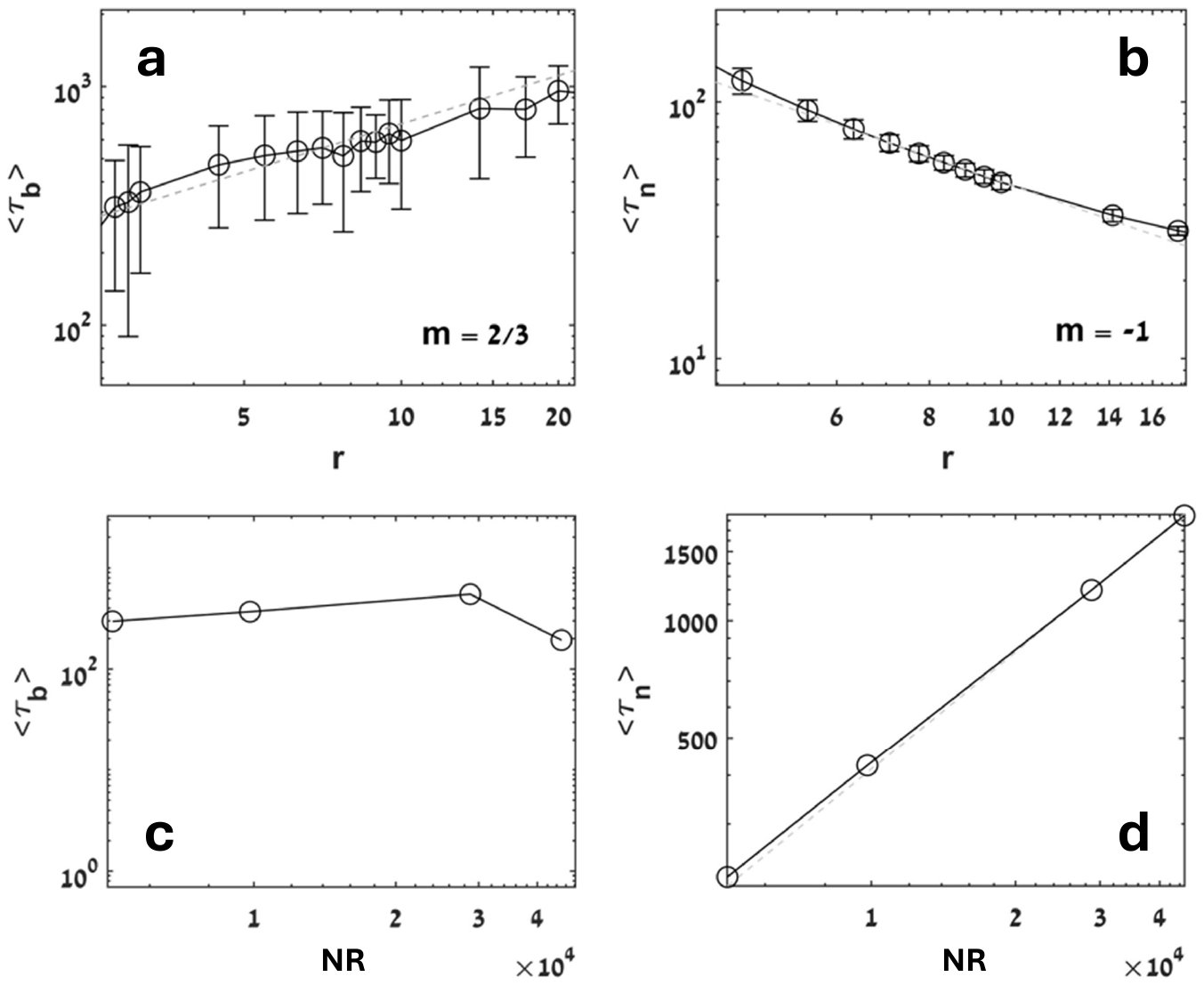}
\caption[Exponents of the nucleation and branching regime duration in SDSN]{\textbf{Exponents of the nucleation and branching (plateau) regime duration as a function of $r$ (SDSN).}
The data in panels (a--d) are averaged over 30 realizations.
\textbf{(a)} Mean duration of the branching (plateau) regime at criticality, as a function of the thermal interaction range $r$. The plateau duration increases with $r$, following a power-law scaling with exponent $m = 2/3$.
\textbf{(b)} Mean duration of the nucleation regime at criticality, as a function of $r$. The nucleation duration decreases with increasing $r$, consistent with a power-law exponent $m = -1$.
\textbf{(c)} For $r = \sqrt3$, the mean duration of the branching (plateau) regime at criticality as a function of the number of segments $NR = 2\times L\times L$. The plateau duration is nearly independent of system size, indicating that it is governed by a microscopic timescale associated with the initiation of nucleation.
\textbf{(d)} For $r = \sqrt3$, the mean duration of the nucleation regime at criticality as a function of $NR$. The nucleation duration increases linearly with system size (exponent $m = 1$).}
    \label{fig:SI_exponents_1Net}
\end{figure*}

Figure~\ref{fig:SI_cooling_1Net} shows the full set of cooling curves for the single self-dependent network, and Figure~\ref{fig:SI_hysteresis_1Net} the corresponding heating--cooling hysteresis; the dynamical exponents extracted from these runs are summarized in Figure~\ref{fig:SI_exponents_1Net}.

\end{document}